\documentclass{aa}  

\usepackage{graphicx}
\usepackage{txfonts}
\usepackage{mathtools}
\usepackage{enumitem}
\usepackage{tikz,hyperref}
\usepackage{multirow}
\usepackage{siunitx}
\usepackage{booktabs}
\usepackage{bm}
\usepackage{float}
\usepackage[normalem]{ulem}

\definecolor{lime}{HTML}{A6CE39}
\DeclareRobustCommand{\orcidicon}{%
    \begin{tikzpicture}
    \draw[lime, fill=lime] (0,0) 
    circle [radius=0.16] 
    node[white] {{\fontfamily{qag}\selectfont \tiny ID}};
    \draw[white, fill=white] (-0.0625,0.095) 
    circle [radius=0.007];
    \end{tikzpicture}
    \hspace{-2mm}
}

\newcommand{\orcidKeerthana}{\href{https://orcid.org/0009-0001-9879-2119}{\orcidicon}}
\newcommand{\orcidEvelyn}{\href{https://orcid.org/0000-0002-2368-6469}{\orcidicon}}
\newcommand{\orcidBoris}{\href{https://orcid.org/0000-0002-1857-2088}{\orcidicon}}
\newcommand{\orcidAugusto}{\href{https://orcid.org/0000-0003-3575-8316}{\orcidicon}}
\newcommand{\orcidRogerio}{\href{https://orcid.org/0000-0002-1321-1320}{\orcidicon}}
\newcommand{\orcidAna}{\href{https://orcid.org/0000-0003-3220-0165}{\orcidicon}}

\begin{document}

   \title{Dual-component stellar assembly histories in local elliptical galaxies via MUSE}

   \author{Keerthana Jegatheesan
          \inst{1}\fnmsep\thanks{\email{ keerthana.jegatheesan@mail.udp.cl}}\orcidKeerthana
          \and
          Evelyn J. Johnston\inst{1} \orcidEvelyn
          \and
          Boris Häu\ss ler\inst{2} \orcidBoris
          \and 
          Augusto E. Lassen\inst{3,4} \orcidAugusto
          \and 
          Rogério Riffel\inst{3}
          \orcidRogerio
          \and 
          Ana L. Chies-Santos\inst{3} \orcidAna}

     \institute{
        Instituto de Estudios Astrof\'isicos, Facultad de Ingenier\'ia y Ciencias, Universidad Diego Portales, Av. Ej\'ercito Libertador 441, Santiago, Chile  
        \and 
        European Southern Observatory, Alonso de Cordova 3107, Casilla 19001, Santiago, Chile 
        \and 
        Universidade Federal do Rio Grande do Sul, Departamento de Astronomia, Av. Bento Gonçalves 9500, Porto Alegre, RS, Brazil
        \and
        INAF-Osservatorio Astronomico di Padova, vicolo dell'Osservatorio 5, I-35122 Padova, Italy
        }

   \date{Received ---; accepted ---}

\abstract
{Elliptical galaxies often exhibit complex assembly histories, and are presumed to typically form through a combination of rapid, early star formation and the subsequent accretion of material, often resulting from mergers with other galaxies. To investigate theories of spheroidal galaxy formation, the objective of this work is to analyse the star formation histories (SFHs) of a sample of three isolated elliptical galaxies in the local Universe observed with MUSE at $z<0.06$. With \texttt{BUDDI}, we decompose the integral field unit (IFU) datacubes into two components with Sérsic profiles, which roughly correspond to the two phases of in situ and ex situ star formation. To constrain the mode of growth in these galaxies, we derived the mass and light-weighted stellar ages and metallicities, and created 2D stellar population maps of each component using \texttt{pPXF}. We reconstructed the mass and light-weighted SFHs to constrain the contribution of different stellar populations to the mass and luminosity of the components through cosmic time. Our results show that the ellipticals in this sample have experienced an early and rapid phase of star formation, either through a rapid dissipative collapse or gas-rich major mergers concentrated in the inner component, which contributes to $\sim50$\% of the galaxy stellar mass. The co-dominant outer component, however, had assembled the bulk of its stellar mass shortly after the inner component did, through accretion via dry mergers and possible gas accretion. This premise is supported by our observations of the inner component being primarily composed of old and metal-rich stars. The outer component has a combination of old and intermediate-age stars, with a moderate spread in metallicities. These results are analysed through the lens of the two-phase scenario, a framework developed over the years to explain the formation histories of elliptical galaxies.
}


   \maketitle
%

\section{Introduction}
\label{sec:intro}

Elliptical galaxies are early-type galaxies (ETGs) that are characterised by a spheroidal appearance and a seemingly featureless and plain morphology that lacks any intricate structure. In the local Universe, they are thought to be simple and massive entities that are `red and dead', hosting old and quiescent stellar populations and meagre amounts of gas to support star formation. However, in reality, this morphology encompasses a wide range of stellar masses and physical sizes, from compact and dwarf ellipticals to the most massive giant ellipticals. Each type has distinct structural properties and surface brightness profiles, providing valuable insights into galaxy formation and evolution. The formation scenarios for elliptical galaxies highlight the complex interplay of physical processes, from rapid monolithic collapse to gradual hierarchical merging and accretion. For instance, an often-cited mechanism for the formation of massive ellipticals is gas-rich mergers between spiral galaxies \citep{toomre1977, Hopkins2008}. These processes essentially shape the structure of the elliptical galaxies, and result in different stellar populations, stellar orbits, and kinematics. Modelling the surface brightness profiles of ellipticals is therefore a crucial step in uncovering their origins and evolutionary history \citep{macarthur2008, vulcani2014, spavone2017, spavone2021}.

Elliptical galaxies were traditionally modelled with a de Vaucouleurs profile \citep{devaucouleurs1948}, which is an $R^{1/4}$ law. This implies a galaxy with a central concentration that becomes more extended radially outward. While this captures the light profile of many ellipticals, improved instrumentation and spatial resolution over the years have revealed a large diversity of elliptical galaxies deviating from this law \citep{Kormendy1977ApJ, king1978}. The Sérsic profile (an $R^{1/n}$ law) was later introduced \citep{sersic1968}, which modelled the surface brightness of the ellipticals more effectively, especially the inner regions of the galaxies. This was possible due to the freedom allowed for the Sérsic index, $n$, which reverts to the de Vaucouleurs profile at $n=4$. This flexibility that is incorporated into the Sérsic law is crucial because although ellipticals are indeed known for their high central concentration, the degree of concentration can vary from galaxy to galaxy. However, even the Sérsic profile might not fully represent the complex substructure that ellipticals are now expected to exhibit \citep{huang2013b,huang2013a,lacerna2016}, challenging the simple structure that is often assumed.

Deep imaging has allowed for the observation of internal substructures in elliptical galaxies, including low-surface-brightness tidal features \citep{ vandokkum2005, tal2009, lazar2023} stemming from their outskirts in cluster environments, or shells \citep{malin1980, athanassoula1985, colbert2001, bilek2016} around the galaxies. Substructures in the central regions of ellipticals have also been observed at different redshifts, such as embedded nuclear discs \citep{graham2016, dullo2018, lacerna2020, deeley2023}. Furthermore, ellipticals have been systematically observed to fall under one of two classes based on the steepness of their central surface brightness profiles: `core ellipticals' with nearly flat central light profiles and `cuspy ellipticals' with sharply peaked central light profiles. The core and cusp regions are considered to be photometrically distinct components of ellipticals \citep{huang2013b}. This central component therefore deviates from the standard Sérsic law that attempts to describe the surface brightness profile of the entire elliptical galaxy with a single model, without accounting for the inner component. It is therefore important to decompose the components of deceptively simple elliptical galaxies to discern their physical relevance, their formation histories, and stellar populations. For instance, \citet{huang2013b} found that their sample of 94 nearby elliptical galaxies in the Carnegie-Irvine Galaxy Survey (CGS) required three or even four components to optimise the modelling of the surface brightness profile. Apart from the structural analysis, a complementary study in \citet{huang2013a} revealed the underlying physical implications of modelling multiple components. Similarly, \citet{lacerna2016} found that their sample of 89 local elliptical galaxies at $z<0.08$ was best modelled with two to four components. More recently, \citet{lima-neto2020} have demonstrated that two or three components are required to model the ellipticals in the NGC 4104 fossil group. The mechanisms leading to the formation of each component are more complex than can be explained by one of the two simple scenarios involving dissipational processes, as was previously mentioned.

The two major pillars of elliptical galaxy formation scenarios are rapid dissipative collapse and hierarchical clustering. Over the decades, both numerical simulations and observations have been able to satisfactorily explain the structure of elliptical galaxies based on these theories. In the monolithic collapse scenario \citep{larson1974, chiosi2002}, an elliptical galaxy is formed by the rapid gravitational collapse of a primordial gas cloud, followed by a substantial dissipation of energy. This theory predicts that ellipticals emerge at high redshifts ($z>5$) directly from the gas, with in situ stellar mass build-up, rather than from the accretion of stars formed ex situ \citep{ogando2005}. The rapid monolithic collapse scenario has been corroborated by strong negative metallicity gradients observed in elliptical galaxies \citep{dominguez-sanchez2019, parikh2019, goddard2017, riffel2023}, owing to the fact that the metals created during supernova explosions are retained by a strong-enough gravitational potential that pushes the metal-enhanced gas towards the centre of the galaxy, where the new stars form at a higher metallicity compared to the outer regions of the galaxy. The hierarchical clustering scenario \citep{toomre1972, cole2000, hopkins2009, avila-reese2014}, on the other hand, predicts that the present-day galaxies formed through major and minor mergers and accretion of galaxies over time. This theory has also been supported by simulations and observations, where a flat or shallow metallicity gradient was exhibited by ellipticals \citep{gonzalez-delgado2015, taylor2017, benedetti2023}, instead of the steep negative gradient expected from the monolithic collapse. Pure merger events are presumed to obliterate traces of the previous stellar history of the galaxy, including metallicity gradients.

To better explain the mixed outcomes of elliptical galaxy observations, a hybrid scenario was proposed \citep{kormendy1989}, according to which both mechanisms play a role in the formation of these remarkably complex galaxies. In the last decades, a two-phase scenario has been gaining traction, challenging the classical pictures of elliptical galaxy formation from either mergers or dissipative gas collapse \citep{kormendy1989}. The two-phase scenario \citep{oser2010, johansson2012, huang2013b, huang2013a} suggests that the first phase of dissipational processes dominate at high redshift through cold accretion and wet mergers of gas-rich galaxies, and that later, at lower redshifts, the galactic evolution is driven by delayed non-dissipational dry mergers. The presence of `red nuggets' at high redshifts, $z\geq1.5$, detected in several studies \citep{cimatti2004, daddi2005, trujillo2007, toft2007, vandokkum2008, damjanov2009} already indicates an early onset of rapid star formation that settles into a compact core. Moreover, a smaller population has been identified of red nuggets that have survived to the present day, having evolved in isolation without perturbations or interactions \citep{ferre-mateu2017, siudek2023, micheli2024, spiniello2024}. Most of these red nuggets eventually accumulate more stellar mass by accreting and merging with smaller systems to acquire the structure that is observed in present-day elliptical galaxies \citep{miller1980, naab2006, oser2010, oser2012}, while some ellipticals seem to have preserved their compact high-$z$ progenitor \citep{barbosa2021}.

Given the intricate structure of elliptical galaxies as both observations and numerical simulations clearly support, we delve into the multi-component context of these galaxies in this paper. We make use of integral field spectroscopy (IFS) with the Multi Unit Spectroscopic Explorer (MUSE) observations of three elliptical galaxies in the local Universe, and perform 2D spectro-photometric decomposition to retrieve their structural parameters, as well as to extract their spectra and stellar populations to interpret their physical significance. The decomposition of a sample of three galaxies serves as a pilot study for further exploring the diverse properties that are inherent to elliptical galaxies over a range of stellar masses, redshifts, and environments.

This paper is organised as follows. Section \ref{sec:data} outlines the data and instrument used in the IFS observations for this study. Section \ref{sec:methods} describes the methods of galaxy decomposition with \texttt{BUDDI}, full spectral fitting with \texttt{pPXF}, and Voronoi binning with \texttt{Vorbin}. The results of this work are highlighted in Sect. \ref{sec:results} in the context of structural parameters, stellar population analysis, and the star formation histories (SFHs) of each component. Section \ref{sec:discussion} provides a discussion on the physical implications and future directions of this study, and we draw our conclusions in Sect. \ref{sec:conclusions}. Throughout this work, we adopt a flat $\Lambda$CDM cosmology with $H_0 = 67.8 \; \mathrm{kms^{-1}Mpc^{-1}, \; \Omega_m = 0.308, \; \Omega_{\Lambda} = 0.692}$ \citep{Planck2016}.

\section{Data and observations}
\label{sec:data}

\begin{table*}[]
\centering
\caption{Co-ordinates and properties of the three elliptical galaxies observed with MUSE.}
\begin{tabular}{@{}ccccc@{}}
\hline \hline
\textbf{Galaxy}               & \begin{tabular}[c]{@{}c@{}}\textbf{RA}\\ (J2000)\end{tabular} & \begin{tabular}[c]{@{}c@{}}\textbf{DEC}\\ (J2000)\end{tabular} & \textbf{\textit{z}} & $\bm{\mathrm{log}(M_*/M_\odot)}$   \\ \midrule
J020536.18-081443.23 & 02:05:36.18                                          & -08:14:43.23                                          & 0.0411 & 10.956\\
J205050.78-004350.85 & 20:50:50.78                                          & -00:43:50.85                                          & 0.0571 & 11.277\\
J225546.96-085457.87 & 22:55:46.96                                          & -08:54:57.84                                          & 0.0590  & 11.274 \\ \bottomrule \\
\end{tabular}

\label{tab:observations}
\end{table*}

MUSE \citep{bacon2010muse} is an optical integral-field spectrograph mounted at the 8.2m ESO Very Large Telescope (VLT) in Chile. It covers the nominal wavelength range of $4600-9350$ Å with spectral sampling of 1.25 Å and a $1\arcmin\!\times\!1\arcmin$ field of view, and spatial resolution of 0.2\arcsec/pixel in the Wide Field Mode (WFM). The MUSE spectrograph has a spectral resolving power ranging from \textit{R}$\simeq$1770 at 4800\,\AA\ to \textit{R}$\simeq$3590 at 9300\,\AA. The three ellipticals were observed under Program ID 099.B-0411 (PI. Johnston) between July and October 2017 using the wide field mode with no adaptive optics (WFM-NoAO-N). The targets initially proposed for observation consisted of 48 galaxies selected from SDSS DR7, representing all major Hubble types from \citet{nair2010}. The selected targets spanned a range of Sérsic indices and B/T ratios, with an inclination $\lesssim40^{\circ}$ derived from \citet{simard2011}, which allowed us to clearly discern the components of the galaxies. Moreover, the sample was chosen to only include unbarred galaxies, with no signatures of recent interactions, such that each galaxy can be cleanly decomposed into its major components. However, being a filler programme observed under sub-optimal weather conditions, only a small set of five elliptical galaxies with a higher set priority were observed. From this, we consider only three galaxies to be suitable for analysis in this work, as the other two ellipticals were severely affected by the observing conditions, making their surface brightness profiles impossible to model. These galaxies were observed with between $3-6$ dithered and rotated exposures, each with an exposure time of 960 seconds, under bright moon conditions, with typical lunar illuminations of $> 85\%$. The sample analysed in this work consists of relatively isolated elliptical galaxies at $z<0.06$, with stellar masses of $\sim10^{11}M_\odot$ from the Nasa Sloan Atlas (NSA) \footnote{Available from \url{http://www.nsatlas.org}}.

The data reduction was carried out using the ESO MUSE pipeline \citep[v2.6,][]{weilbacher2020} in the ESO Recipe Execution Tool (\texttt{EsoRex}) environment \citep{esorex}. The master bias, flat field, and wavelength calibrations for each CCD were created from the associated raw calibrations, and were applied to the raw science and standard-star observations as part of the pre-processing steps. Flux calibration was carried out using the standard star observations from the same nights as the science data, with the sky continuum measured directly from the science exposures with subsequent background subtraction. The individual exposures were aligned to a common frame and stacked to produce the final datacube. To alleviate residual sky contamination, we used the Zurich Atmosphere Purge code \citep[\texttt{ZAP}, ][]{soto2016}. The co-ordinates and redshifts of the galaxies used in this analysis are listed in Table \ref{tab:observations}. For readability, these galaxies will henceforth be denoted as J020536 (MRK 1172), J205050, and J225546.

\section{Methods}
\label{sec:methods}

In this section, we outline our methodology for decomposing the spectra of the elliptical galaxies using the BUlge-Disc Decomposition of integral field unit (IFU) datacubes (\texttt{BUDDI}) software, followed by fitting the resulting spectra with \texttt{pPXF}. Additionally, we detail the construction of 2D stellar population maps utilising the Voronoi tessellation technique with \texttt{Vorbin}.

\subsection{Galaxy decomposition with BUDDI}
\label{subsec:buddi}

The light profiles of the elliptical galaxies in this sample were modelled using \texttt{BUDDI} \citep{johnston2017buddi}, which decomposes galaxies in IFU datacubes into multiple components and extracts each spectrum separately. This code functions as an IDL wrapper for \texttt{GalfitM} \citep{vika2013, haeussler2013megamorph, vika2014, haeussler2022galfitm}, extending its capabilities to handle the vast spectral dimension of IFU datacubes. While it would be ideal to freely model each image of a datacube with $\sim4000$ slices, this endeavour would require immense amounts of computational time and memory, which doubles with every additional component that is included in the model. \texttt{BUDDI} works around this limitation by first establishing the structure of the galaxy in a series of steps, then fitting only the magnitudes at each wavelength slice to construct the component spectra. This methodology enables clean separation of component spectra that would otherwise be contaminated by overlapping light from different galaxy components and, in some cases, the sky background (as is demonstrated later in this section).

The galaxies were each modelled at first with a single Sérsic profile, and the complexity in the model was introduced by subsequent addition of a second Sérsic component. A double Sérsic model with an additional Point Spread Function (PSF) component, and a triple Sérsic model were also tested, but they either did not yield meaningful results or failed to fit the galaxy (we refer the reader to Sect. \ref{subsec: models_params} for details on the preferred choice of components). Therefore, throughout this study, we adopt the models based on single and double Sérsic light profiles. We refer the reader to \citet{johnston2017buddi} for the full details of the IFU decomposition with \texttt{BUDDI}, and present a brief overview here.

\texttt{BUDDI} requires some initial data preparation before starting the fitting process. This includes normalising the galaxy kinematics by measuring the line-of-sight (LOS) velocities ($V$) and velocity dispersions ($\sigma$), and introducing the following kinematics corrections: (i) broadening the spectrum of each spaxel to match the maximum LOS velocity dispersion, and (ii) shifting the spectrum to match the LOS velocity at the centre of the galaxy. The kinematics corrections eliminate any artefacts in the final spectral features induced by the variation in rotation and velocity dispersion across the galaxy. For this, the IFU datacube was first binned using the Voronoi tessellation technique implemented by the \texttt{Vorbin} Python package\footnote{Available from \url{http://purl.org/cappellari/software}} package of \citet{cappellari&copin2003}, with a target minimum S/N = 50 $Å^{-1}$.
Subsequently, the stellar kinematics were measured by full spectral fitting of the binned spectra using the penalised PiXel Fitting routine (\texttt{pPXF\footnote{See footnote 2}}) of \citet{CappellariEmsellem2004PASP}. We use the single stellar population (SSP) models \citep{vazdekis2010, vazdekis2015sps} based on the Medium resolution INT Library of Empirical Spectra  \citep[MILES; ][]{sanchez-blazquez2006miles}, constructed from the PADOVA2000 stellar evolutionary tracks \citep{padovamodels}. The parameter space for the kinematics measurements spans stellar ages between 1 and 17.78 Gyr, and metallicities between -1.71 and 0.22, resulting in a total of 156 template spectra.

The setup for \texttt{BUDDI} requires the following input files:
\begin{enumerate}[label=(\roman*)]
    \item The flux datacube, which has been log-rebinned in wavelength.
    \item The PSF datacube, which has also been log-rebinned in wavelength. In all three datacubes, postage stamps of stars in the field of view were created and stacked at each wavelength. A Gaussian profile was used to model them, constructing the PSF datacube. The full widths at half maximum of the PSFs in our sample range from $0.8\arcsec-1.4\arcsec$ at the central $r-$band wavelength of 6166 Å.
    \item The bad pixel mask datacube, which identifies and masks the pixels with no valid flux values. This also includes the masking of any bright sources in the field of view that can affect the fitting of the light profile of the target galaxy. 
    \item The sigma datacube, which is crucial for accurately determining the flux uncertainty in each individual pixel.
\end{enumerate}

Once the datacubes are in place, the fitting process in \texttt{BUDDI} takes place in three major steps. We refer the reader to \citet{johnston2017buddi} and \citet{johnston2022buddi1} for details.

\begin{enumerate}
    \item Fitting the median `broad-band' image. \\
    The datacube was collapsed in wavelength to a median-stacked white-light image. Within the \texttt{BUDDI} routine, this single image was fitted using \texttt{GalfitM} with a single Sérsic component at first, and subsequently with two Sérsic components. The sky background amplitude and the gradient in the $x$ and $y$ directions were allowed to be freely fitted simultaneously - this was deemed necessary due to the inconducive observing conditions, which had created a non-uniform sky background. \\
    
    \item Fitting the `narrow-band' images. \\
    This step fine-tunes the Chebyshev polynomials in \texttt{GalfitM}, which model the dependence of structural parameters ($R_e$, $n$, $b/a$, PA) on wavelength for the galaxy in the single Sérsic model, and for both components in the double Sérsic model. The datacube was binned along wavelength into 10 narrow-band images \citep{johnston2022buddi1, johnston2022buddi2, jegatheesan2024} and fitted simultaneously twice: first with complete freedom for the parameters (10th-order Chebyshev polynomial), and then manually choosing the appropriate polynomial order after visual inspection of the intermediate plots that \texttt{BUDDI} creates with the estimated structural parameters as a function of the wavelength. For each galaxy and component, the chosen polynomial orders for the parameters are listed below in Table \ref{tab:polynomials}: a first-order polynomial constrains the parameters to remain constant with wavelength, while a second-order polynomial allows for linear variation in wavelength. The magnitudes were allowed to have complete freedom during the fit, and the initial parameters were set to those measured in each preceding step. \\

\begin{table}[h]
\centering
\caption{Chebyshev polynomials introduced in the step fitting the narrow-band images of the datacube.}
\begin{tabular}{@{}cccccc@{}}
\hline \hline
\textbf{Galaxy}                   & \textbf{Comp} & $\bm{R_e}$ & \textbf{$\bm{n_{ser}}$} & $\bm{b/a}$ & \textbf{PA} \\ \midrule
\multirow{2}{*}{\textbf{J020536}} & single          & 2              & 2            & 1            & 1           \\

& inner              & 1              & 2            & 1            & 1           \\
                                  & outer              & 1              & 2            & 1            & 1           \\ \midrule
\multirow{2}{*}{\textbf{J205050}} & single          & 2              & 2            & 1            & 1           \\
& inner              & 1              & 1            & 1            & 1           \\
                                  & outer              & 1              & 1            & 1            & 1           \\ \midrule
\multirow{2}{*}{\textbf{J225546}} & single          & 1              & 2            & 1            & 1           \\
& inner              & 1              & 1            & 1            & 1           \\
                                  & outer              & 1              & 2            & 1            & 1           \\ \midrule \\
\end{tabular}

\label{tab:polynomials}
\end{table}

    \item Fitting each image slice of the datacube.\\
    The final step makes use of the refined parameters from the previous narrow-band fits and keeps them fixed for each image slice, while allowing only the magnitudes to vary in the fit. This estimates the magnitude, and therefore the flux of each component, in each image slice of the datacube, ultimately extracting their clean 1D spectra.
    
\end{enumerate}

\begin{figure*}
    \begin{center}
        \includegraphics[trim=1.8cm 0cm 1.8cm 1cm, clip, width=\textwidth]{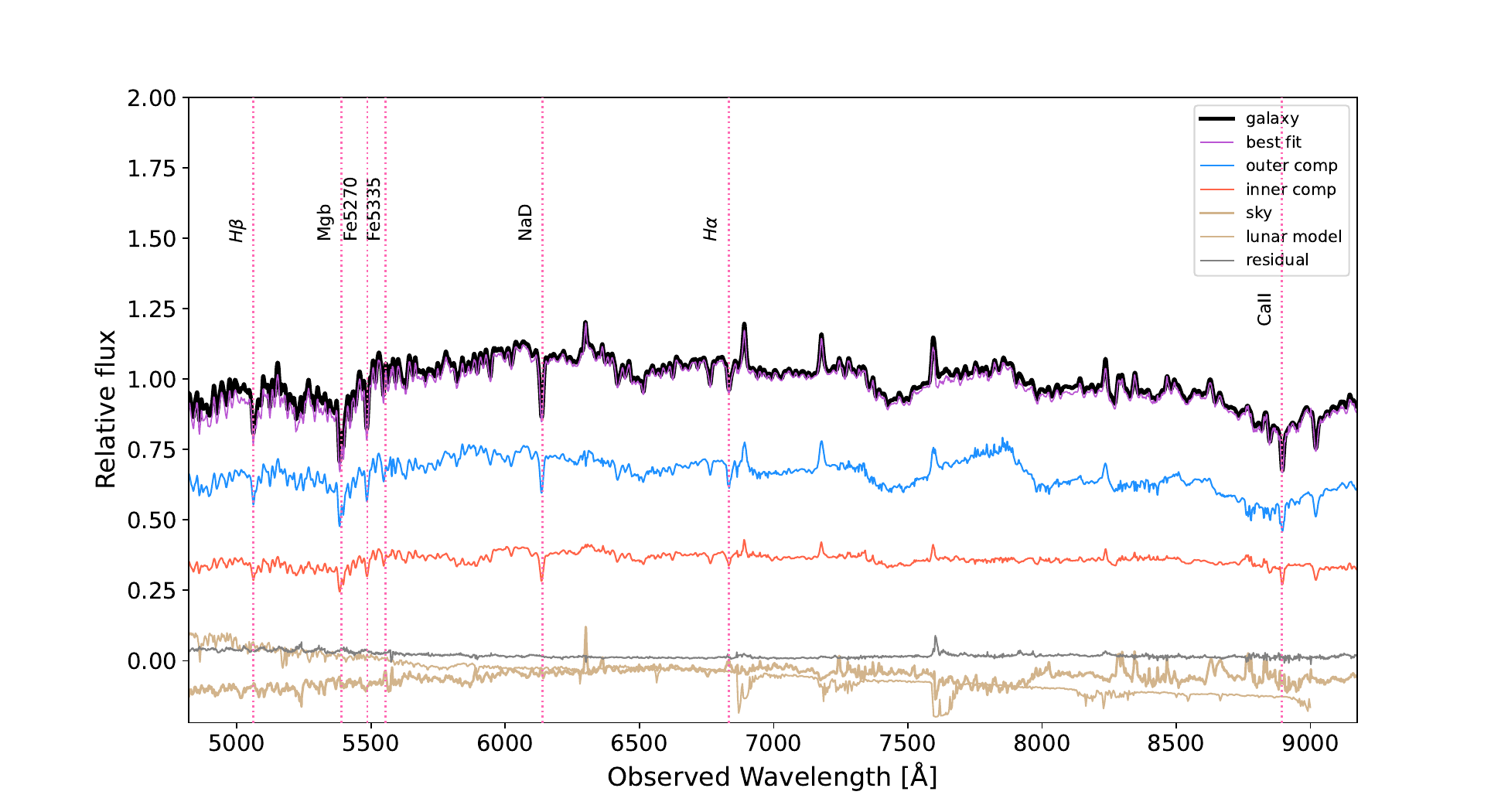}
        \caption{Decomposed spectra of the different components in J020536 extracted by \texttt{BUDDI}, for the MUSE wavelength range from the optical to near-infrared. The galaxy spectrum integrated directly from the datacube is shown in black. The spectra of the inner component and outer component are displayed in red and blue, respectively, while the thick brown curve represents the sky spectrum from \texttt{BUDDI}. The scattered moonlight modelled by \texttt{SkyCalc} is shown by the thin brown curve, which has been arbitrarily scaled to match the flux scale of the normalised galaxy spectrum. The best fit model is represented by the purple spectrum, which includes the contributions of the inner and outer components, as well as the sky component. The residual flux of the best fit spectrum to the observed galaxy spectrum is displayed in grey. The dotted pink lines mark the wavelength positions of several important spectral features that are indicators of stellar age and metallicity.}
        \label{fig:decomposed_J020536}
        
    \end{center}
\end{figure*}

Having modelled the surface brightness of the galaxies with one, two, and three Sérsic profiles, we opted for the two-component models for all of them, which are shown in Fig. \ref{fig:models}. To unveil the physical properties and implications associated with the different components of elliptical galaxies, the clean spectrum of each component was extracted with \texttt{BUDDI}. The magnitudes from \texttt{GalfitM} correspond to the total fluxes of each Sérsic profile integrated out to infinity. This extends to the individual image slice models in the datacube, and is reflected in the final 1D spectrum derived by \texttt{BUDDI}, which was obtained by plotting these fluxes as a function of wavelength \citep{johnston2017buddi}. The decomposed spectra of the inner and outer components, the sky, and that corresponding to the best fit model (inner + outer + sky) are shown for J020536 in Fig. \ref{fig:decomposed_J020536}, and in Figs. \ref{fig:decomposed_J205050} and \ref{fig:decomposed_J225546} for J205050 and J225546, respectively. We note that since most of the observing nights for the filler programme included high lunar illumination (Sect. \ref{sec:data}), we have included an arbitrarily scaled scattered lunar model spectrum for each galaxy, obtained from the ESO \texttt{SkyCalc} \footnote{Available from \url{https://www.eso.org/observing/etc/bin/gen/form?INS.MODE=swspectr+INS.NAME=SKYCALC}} Sky Model Calculator tool \citep{noll2012, jones2013}. The effect of scattered moonlight is strong for galaxies J205050 and J225546, and closely matches the slope of the sky spectrum modelled by \texttt{BUDDI}, effectively demonstrating its ability to cleanly disentangle the light from physical galaxy components and from sky contributions. For J020536, the scattered moonlight has less of a contribution to the modelled sky spectrum. However, since the MUSE datacubes undergo two sky subtraction steps (once within the \texttt{EsoRex} pipeline, and later enhanced with the \texttt{ZAP} code), the lunar contribution could in principle be partially eliminated, leading to the slope differences between the scattered moonlight spectrum and the sky spectrum, especially in J020536. An important point to note is that wherever there is a known telluric feature in the lunar spectrum (for instance, the O2 A-band at $\sim7600$ Å in absorption), there appears a spectral feature in the sky spectrum modelled by BUDDI. The spectral feature in the sky spectrum may or not be in absorption due to the uncertainties in sky subtraction for the MUSE cube (including possible over-subtraction in some cases); it is nevertheless encouraging that BUDDI picks up these features at the expected wavelength. Software is often only as effective as the quality of the data it processes - even so, it has clearly demonstrated its effectiveness in cleanly decomposing the component spectra from the sky spectrum. The decomposed spectra of the components were then used for the stellar population analyses, in order to unveil the physical properties and implications associated with the different components of elliptical galaxies.

\subsection{Spectral fitting for stellar population analysis}
\label{subsec:ppxf}


The spectra of the two components obtained from the previous step (Sect. \ref{subsec:buddi}) were fitted using \texttt{pPXF} to derive the stellar metallicity [M/H] and stellar age. The Python implementation of \texttt{pPXF} v9.1.1 was used in this work. \texttt{pPXF} finds a linear combination of stellar templates that best matches the input spectrum. For this study, both the mass-weighted and light-weighted stellar population properties were derived and analysed. The full spectral fitting method optimises the mass or light weights for each SSP spectrum in the template library, such that the observed spectrum is a combination of these weights \citep{cappellari&copin2003, cappellari2017ppxf, cappellari2023}. The mean logarithmic mass and light-weighted stellar ages and metallicities estimated by \texttt{pPXF} are defined as:

\begin{equation}
    \mathrm{\langle log(age)\rangle = \frac{\Sigma\omega_{i} \times log(age_{temp,i})}{\Sigma\omega_{i}}}
,\end{equation}

\begin{equation}
    \mathrm{\langle[M/H]\rangle = \frac{\Sigma\omega_{i} \times [M/H]_{temp,i}}{\Sigma\omega_{i}}}
.\end{equation} 

The empirical library of stellar spectra employed was MILES \citep{vazdekis2015sps}\footnote{Available from \url{http://miles.iac.es/pages/webtools.php}}, along with the BaSTI isochrone models \citep{pietrinferni2004basti}, which track stellar evolution. This choice of the stellar library is attributed to the vast coverage of stellar metallicities and ages, which is necessary for studying nearby galaxies in the local Universe, which can contain both old and young stars with low and high metallicities.  Restricting to the safe ranges defined in \citet{vazdekis2010}, the chosen stellar population parameters span a range from -1.79 to 0.26 dex for the metallicities [M/H] and from 30 Myr to 14 Gyr for the ages, forming a combined total of 530 template spectra. Since each stellar template is normalised to 1$M_{\odot}$, \texttt{pPXF} derives the mass weights by default. In order to estimate the light weights, the stellar templates were normalised by the $V-$band luminosity ($5070-5950$ Å), chosen such that the resulting properties favour more recent star formation episodes in the galaxy. The Universal Kroupa IMF \citep{kroupa&boily2002imf} with a slope of 1.3 is assumed.

During the spectral fit with \texttt{pPXF}, an 8th-order multiplicative Legendre polynomial was used to correct for the shape of the continuum. Furthermore, the inclusion of a multiplicative polynomial in the fit also ensures that the spectral fit is independent of dust reddening, thereby eliminating the need to define a reddening curve explicitly within the \texttt{pPXF} routine \citep{cappellari2017ppxf}. Additionally, the [OI]5577Å sky line was masked during the fit. The code allows for emission lines to be fitted simultaneously alongside the stellar continuum using a set of respective Gaussian templates representing the nebular emission. The spectral range for the fit was constrained to rest-frame $4700-6780$ Å due to inconsistencies that can be introduced when using the full MUSE wavelength range \citep{krajnovic2014}. Moreover, this chosen range spans the important spectral features that are sensitive to stellar age and metallicity ($\mathrm{H\alpha}$, $\mathrm{H\beta}$, $\mathrm{Mg_2}$, and Fe) and aligns with the wavelength coverage in the MILES models.

\begin{figure*} 
    
  \centering
  \begin{tabular}{cc}
    \includegraphics[trim=0.4cm 0cm 0cm 0cm, clip, width=0.49\textwidth]{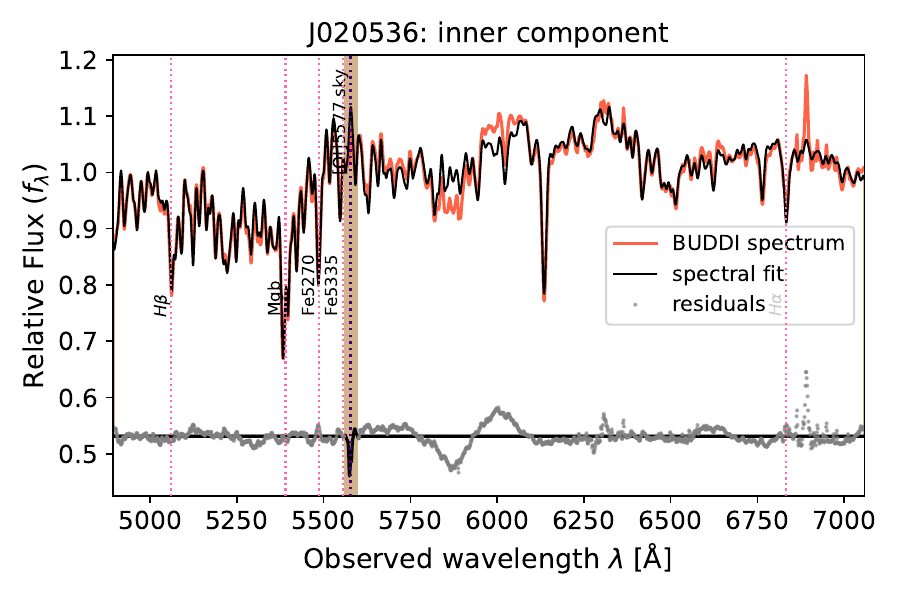} & 
    \includegraphics[trim=0.4cm 0cm 0cm 0cm, clip, width=0.49\textwidth]{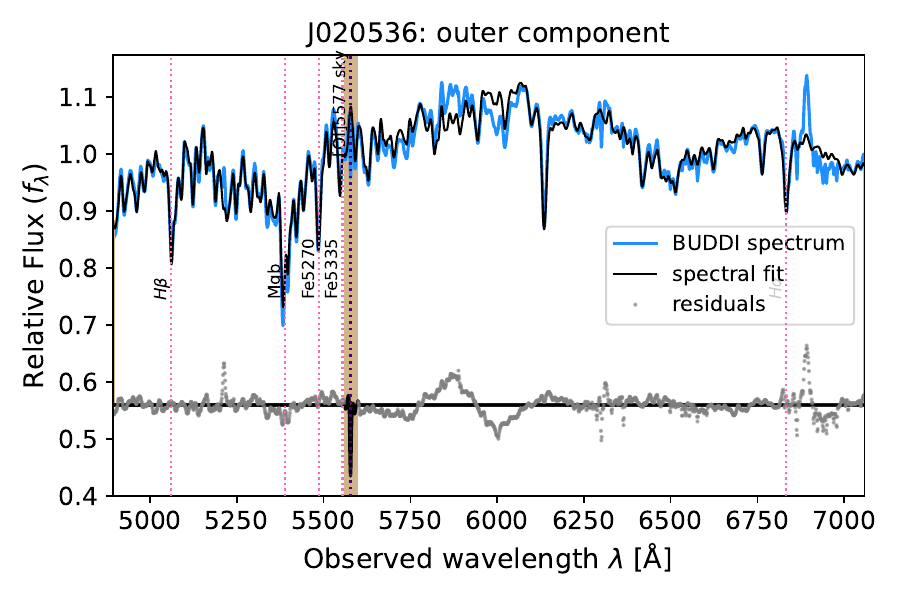} \\ 
    \includegraphics[trim=0.4cm 0cm 0cm 0cm, clip, width=0.49\textwidth]{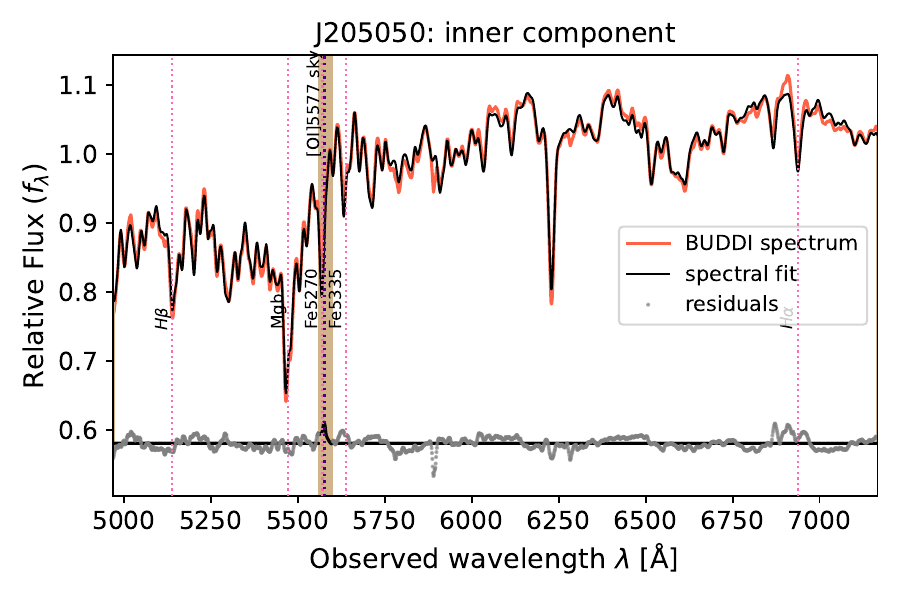} &
    \includegraphics[trim=0.4cm 0cm 0cm 0cm, clip, width=0.49\textwidth]{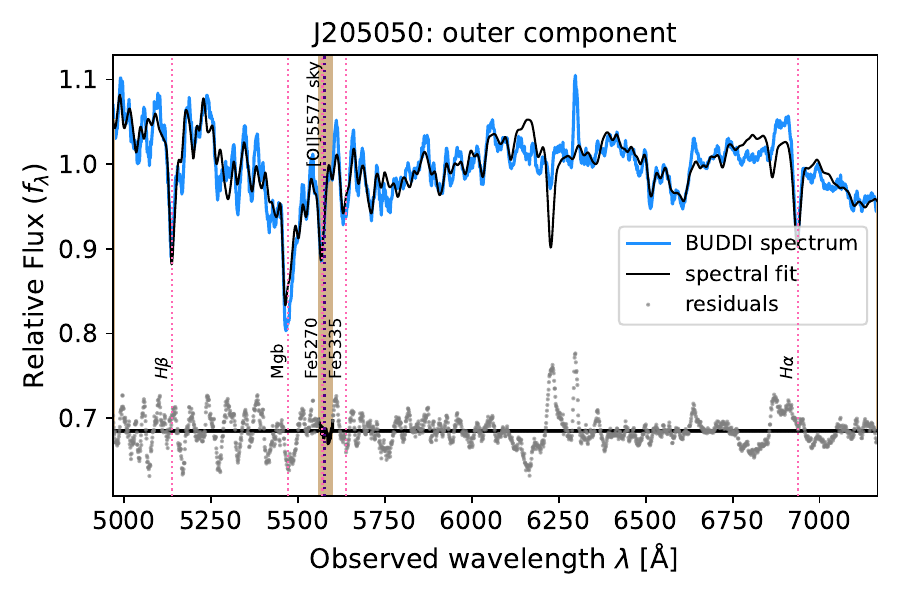} \\ 
    \includegraphics[trim=0.4cm 0cm 0cm 0cm, clip, width=0.49\textwidth]{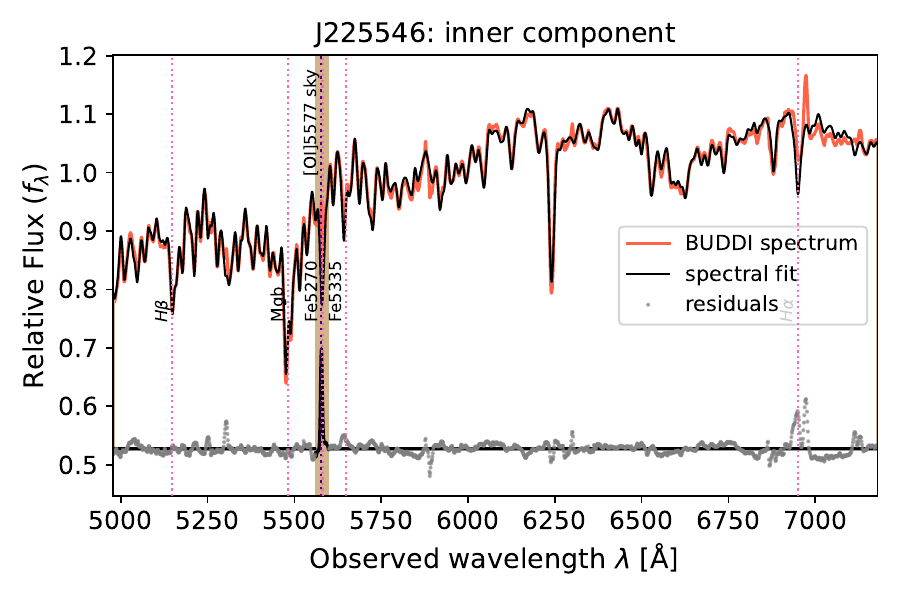} & 
    \includegraphics[trim=0.4cm 0cm 0cm 0cm, clip, width=0.49\textwidth]{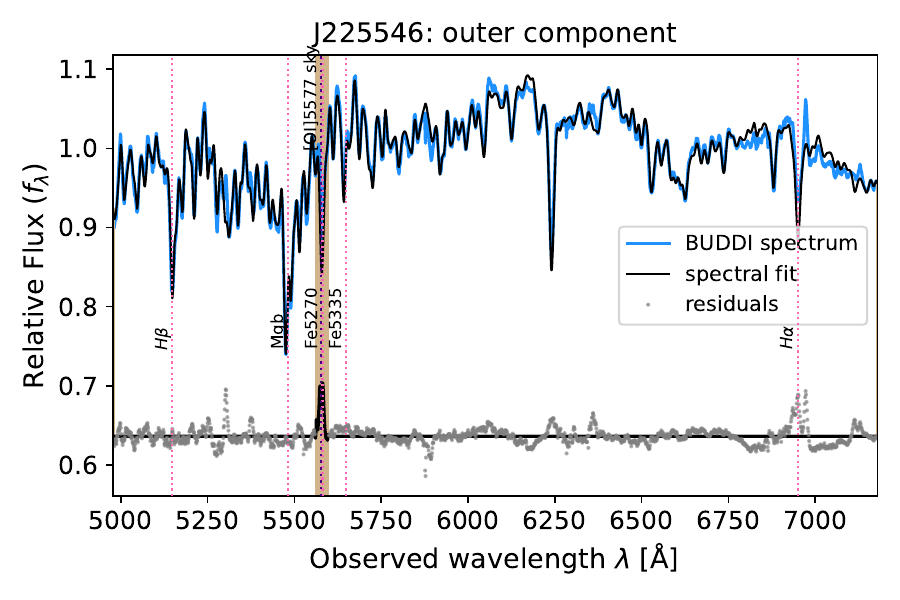} \\ 
    
  \end{tabular}
  \caption{pPXF fits to the inner and outer component spectra derived from \texttt{BUDDI}. The component spectra are shown in red and blue for the inner and outer components, respectively, while the best spectral fits are shown in black. The grey points show the residuals, defined as the difference between the spectrum and the best fit. The brown stripe marks the location of the [OI] sky line, which is masked out during the fits.}
  \label{fig:ppxf_fits}
\end{figure*}

The \texttt{pPXF} routine provides a solution to the ill-posed problem of extracting stellar population properties from observed spectra. The extensive span of the parameter space of the SSPs inevitably leads to non-unique solutions that are degenerate in metallicity or age. This is one of the commonly cited challenges in the fossil-record method with stellar population modelling and full spectral fitting, along with the imprecise age resolution for older populations with $t>8$ Gyr \citep{cid-fernandes2005, riffel2009, ibarra-medel2019, lacerna2020, riffel2024}. The solution incorporated within \texttt{pPXF} is regularisation, which is responsible for mitigating intrinsic degeneracies in the final solution. When a possible degeneracy is encountered, the inclusion of a regularisation term treats this by smearing out the weights of similar metallicities and ages, going from discrete weights that resemble a single stellar population to a solution with smooth weights that are more physically motivated in galaxies. The \texttt{REGUL} parameter within the routine sets the degree of smearing, following the prescription outlined in the \texttt{pPXF} user manual and in \citet{shravan2015} and \citet{jegatheesan2024}. A word of caution is advised at this point while interpreting the SFHs from full spectral fitting: star-formation is a stochastic and unpredictable process, which can occur on timescales shorter than those specified in the synthetic template models. The spectral fits for each component spectrum in the three elliptical galaxies resulting from \texttt{pPXF} are displayed in Fig. \ref{fig:ppxf_fits}. We find substantially robust fits for all objects, with the exception of the outer of component of J205050. From Fig. \ref{fig:decomposed_J205050}, this galaxy appears to have the strongest lunar contamination, which has mostly been modelled off as the sky spectrum. However, if there is a fraction of residual scattered moonlight contaminating especially the outer regions, this can add to the noise of the spectrum of the outer component. This could lead to a less robust fit to the spectrum using \texttt{pPXF}; however, we still find that the crucial spectral features required for stellar population analysis have been modelled fairly well, and though we caution the direct interpretation of the fits and the resulting stellar populations for this component, we note that the SFHs are consistent with the other galaxies.

\subsection{Spectral index measurements for stellar population analysis}

We additionally measured the line strengths of stellar age and metallicity sensitive absorption features in the spectra through the Lick indices defined in \citet{worthey1999}. This analysis was done as a consistency check to empirically determine the stellar populations from the spectra, due to potential model-dependent uncertainties and the effects of trade-offs achieved with the choice of the regularisation parameter, which can be introduced in the age-metallicity distribution estimated through full spectral fitting with pPXF. The absorption line strengths were measured using the public Python library \texttt{PyLick}\footnote{Available from \url{https://gitlab.com/mmoresco/pylick/}} \citep{borghi2022}, which performs a fast estimation of the predominant Lick indices and their uncertainties, spanning the UV to the near-infrared region. The uncertainties were estimated following the method of \citet{cardiel1998}, through simulations accounting for the errors in the LOS velocities and S/N in the observed spectrum. 

The H$\beta$ index served as the spectral diagnostic for stellar age, while the metallicity was determined using the composite index [MgFe]$'$ = $\sqrt{\mathrm{Mg}b \; (0.72\times \mathrm{Fe5270} + 0.28 \times \mathrm{Fe5335})}$, which is less sensitive to the $\alpha$-element abundance and serves as a more robust indicator of the total metallicity of the galaxy or component. These indices were plotted on a H$\beta-$[MgFe]$'$ plane in Fig. \ref{fig:lick_ind}, overlaid with the SSP model grids defined in \citet{vazdekis2010}. The chosen grid spans ages from $2 - 14$ Gyr and metallicities [M/H] from -1.79 to  +0.26 dex, constructed with the Lick index measurements from the MILES stellar library \citep{sanchez-blazquez2006miles}. The flexibility of the MILES webtools lies in its ability to customise the SSP models by convolving the model spectra with the velocity dispersion of the galaxy, effectively accounting for line broadening effects in the observed spectrum. As such, we used a convolved model with a velocity dispersion of 212~$\mathrm{km~s^{-1}}$, which represents the median dispersion of the three elliptical galaxies in the sample. Although the grid is essential for estimating light-weighted stellar population properties, we have used it here solely to validate our \texttt{pPXF} results, in order to verify their reliability in the presence of potential intrinsic uncertainties.

\subsection{Voronoi binning for spatially resolved properties}
\label{subsec:voronoi}

As a complementary analysis, the stellar populations across the spaxels of the datacube were measured and visualised as a 2D map to study the properties of the inner and outer regions of the galaxies. Given the variations in S/N across the spaxels in the MUSE datacube, we opted to employ Voronoi binning, which ensures that a minimum user-defined S/N per bin is achieved throughout the galaxy. The size of the Voronoi bins adapts based on the corresponding S/N, which allows for the spaxels of low S/N to be handled effectively \citep{cappellari&copin2003} by assigning them to larger bins, while high S/N spaxels, usually located in central galaxy regions, are assigned to smaller bins or left unchanged \citep{cappellari&copin2003}.

The Python implementation of the \texttt{Vorbin} package of \citet{cappellari&copin2003}
was used to Voronoi bin the S/N in our galaxy datacubes. The continuum S/N was determined at a rest-frame wavelength of 5635 Å, in a region without any spectral features. According to \citet{zibetti2023}, a minimum spectral S/N of 20 $\AA^{-1}$ is required to constrain stellar ages reliably from stellar population synthesis (SPS) models  \citep[see also][]{cid-fernandes2005, riffel2009}. We therefore chose the constant S/N threshold to be more conservative at 50 $\AA^{-1}$, following \citet{johnston2021muse}, in which it was proven to be sufficient to extract stellar populations. The binning was performed such that the total S/N of each assigned bin was above the set threshold, thereby optimising the usage of spaxels on the outskirts of the galaxy with low S/N in addition to the spaxels with high S/N in the core of the galaxy. Any foreground stars in the MUSE field of view or neighbouring objects were masked out wherever necessary, as well as the spaxels with S/N$<3$ to discard noise-dominated spaxels and sky spaxels. 

The spectra in each bin were then summed, and modelled with \texttt{pPXF}, following the prescription detailed in Sect. \ref{subsec:ppxf}, in order to extract the mean stellar metallicities and stellar ages. We note that this technique provides only a general overview of the stellar population properties of the galaxy in a 2D map, and allows us to study these properties and any features in inner and outer regions of the galaxy. While this is useful as a confirmation analysis, it is not an exact alternative to the IFU galaxy decomposition described in Sect. \ref{subsec:buddi}. The Voronoi-binned maps represent the light of the different components superposed on each other as the total light in each spaxel, and therefore this contamination propagates in each bin. \texttt{BUDDI}, on the other hand, extracts the light from each underlying component of the galaxy, minimising the superposition.

\section{Results}
\label{sec:results}

\subsection{Component models and structural parameters}
\label{subsec: models_params}

\begin{figure}
    \begin{center}
    \includegraphics[trim=1.4cm 0cm 0.3cm 0cm, clip, width=0.98\columnwidth]{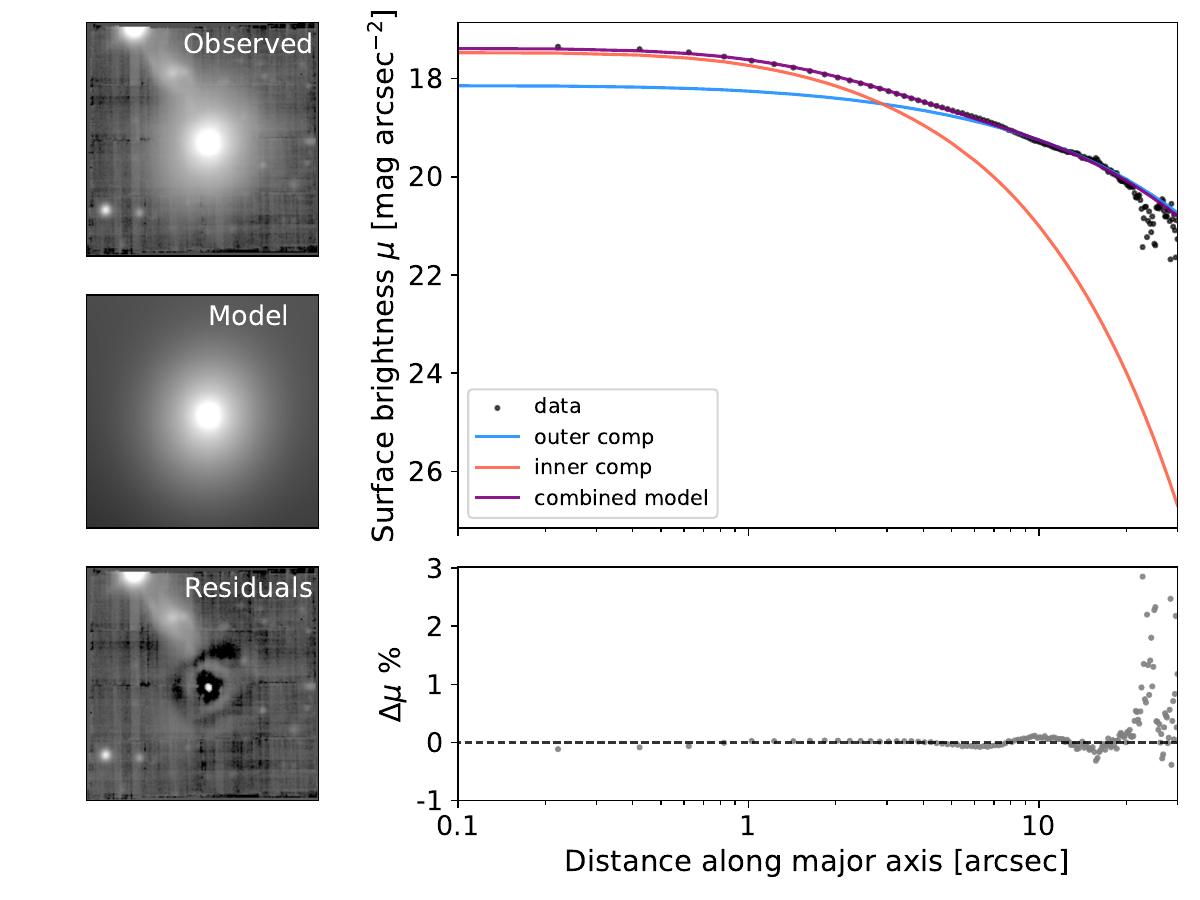}

    \includegraphics[trim=1.4cm 0cm 0.3cm 0cm, clip,width=0.98\columnwidth]{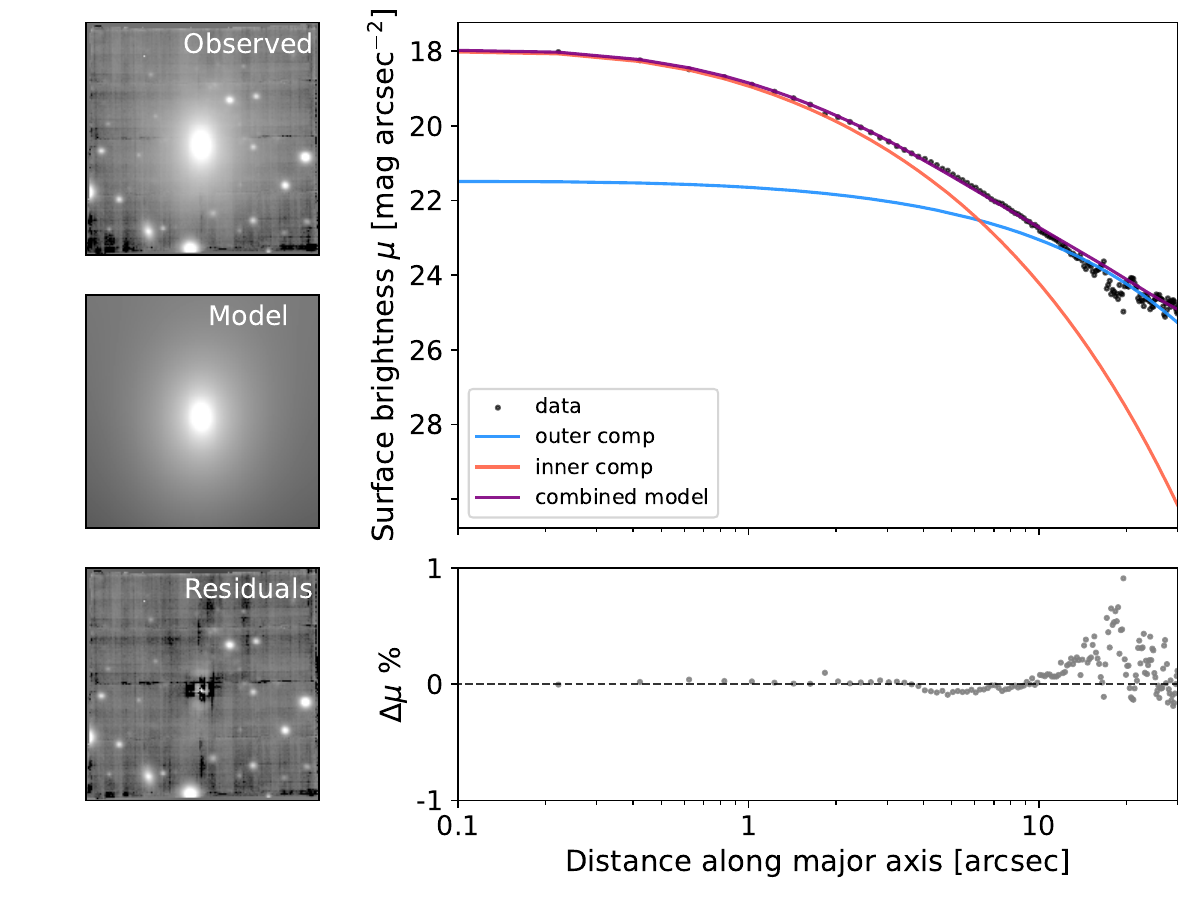}

    \includegraphics[trim=1.4cm 0cm 0.3cm 0cm, clip,width=0.98\columnwidth]{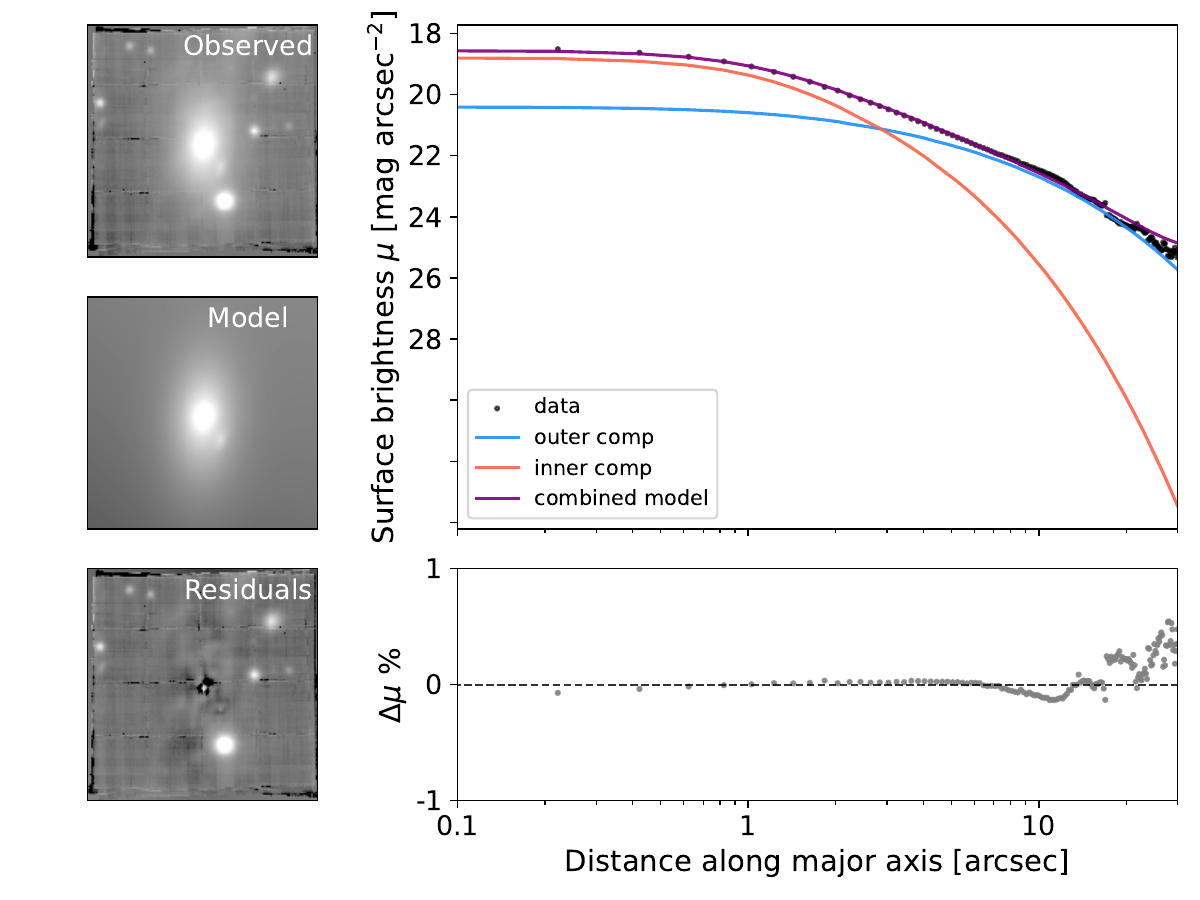}

    \caption{Fit to the median-stacked white-light image of the three ellipticals, J020536, J205050, and J225546, from top to bottom. Left column: Observed input image, best-fit model, and residual image. The images have all been scaled to the same flux for comparison. Upper right panel: 1D light profile of the galaxy for the white-light image along the major axis (black points), the Sérsic profiles of the inner (red line) and outer (blue line) components and the combined model (purple line). Lower right panel: Residuals (in percent) of the 1D data points and model as a function of distance along the major axis. } 
    \label{fig:models}
    \end{center}
\end{figure}

As is described in Sect. \ref{subsec:buddi}, the galaxies were modelled with a single Sérsic profile at first, and then the complexity of the models was subsequently increased and tested using \texttt{BUDDI} in order to isolate any relevant substructure in elliptical galaxies presenting different physical and stellar population properties. We attempted to go as far as three components, including a double Sérsic profile with a PSF model, and a triple Sérsic component model, drawing from the findings in \citet{huang2013b} and \citet{huang2013a}. However, adding a third component caused the fit to crash before completion due to a forced attempt at fitting a non-existent component for two of the galaxies (J205050 and J225546). Adding a PSF component to the double Sérsic model did not cause the fits to fail, but did not contribute to a significant difference to the models and spectra. For J020536, the fit appeared to be successful with three Sérsic components, only considering the residual image and the estimated structural parameters. However, on closer inspection, the fit had in fact resulted in spectra with a distorted and erratic continuum, suggesting inaccuracies due to significant noise. Given the spatial resolution offered by MUSE and the sub-optimal observing conditions, a double Sérsic model was chosen to be the best fit for all three galaxies, which was decided after visual inspection of the fit residuals and the estimated structural parameters, along with the extracted spectra. The structural parameters estimated by \texttt{BUDDI} are listed in Table \ref{tab:params} and the fits to the galaxies in Fig. \ref{fig:models}.

\begin{table}[h!]
\centering
\caption{Structural parameters and flux percentages for the single and two-component models.}
\tabcolsep=0.10cm
\begin{tabular}{@{}cccccccc@{}}
\hline \hline
\textbf{Galaxy}                   & \textbf{Comp} & $ \Biggl (\dfrac{\bm{F_{comp}}}{\bm{F_{total}}}\Biggr)_{\bm{r}}$ & \textbf{\begin{tabular}[c]{@{}c@{}}$\bm{R_e}$\\ (arcsec)\end{tabular}} & \textbf{\begin{tabular}[c]{@{}c@{}}$\bm{R_e}$\\ (kpc)\end{tabular}} & \textbf{$\bm{n_{ser}}$}  & \textbf{$\dfrac{\bm{b}}{\bm{a}}$} & \textbf{\begin{tabular}[c]{@{}c@{}}PA\\ (deg)\end{tabular}} \\ \midrule
\multirow{3}{*}{\textbf{J020536}} & single        &                                          & 4.33    & 3.63                                                    & 2.35               & 0.90                   & -2.26                                                       \\
                                  & inner         & 0.34                                      & 1.68 & 1.41                                                           & 1.21               & 0.95                   & 6.05                                                       \\
                                  & outer         & 0.66                                    & 7.96 & 6.67                                                        & 1.43            & 0.85                   & -8.47                                                        \\ \midrule
\multirow{3}{*}{\textbf{J205050}} & single        &                                          & 5.10       & 5.79                                                 & 2.96               & 0.68                   & 2.48                                                        \\
                                  & inner         & 0.45                                    & 2.49 & 2.84                                                        & 1.95               & 0.68                   & 3.62                                                       \\
                                  & outer         & 0.55                                    & 16.61 & 18.98                                                        & 1.34               & 0.73                   & -4.74                                                        \\ \midrule
\multirow{3}{*}{\textbf{J225546}} & single        &                                          & 5.85      & 6.88                                                  & 2.71               & 0.64                   & -5.41                                                       \\
                                  & inner         & 0.34                                   & 2.05 & 2.42                                                        & 1.66               & 0.78                   & -11.76                                                       \\
                                  & outer         & 0.64                                   & 10.33   & 12.16                                                       & 1.55               & 0.50                   & -2.77                                                      \\ \midrule \\
\end{tabular}
\tablefoot{Structural parameters and the flux contribution of each component relative to the total flux for the single-component model (`single') and for each component of the two-component model (`inner' and `outer') at the $r-$band central wavelength 6166 Å. The conversion of $R_e$ from arcseconds to kiloparsecs was computed with $R_e$ [kpc] = $4.848 \times 10^{-6}$ $\times$ $dA(z)$ [kpc] $\times$ $R_e$ [arcsec], where $dA$ is the redshift-dependent angular diameter distance in kiloparsecs to each galaxy.}

\label{tab:params}
\end{table}

Considering the two-component fits, the Sérsic indices of all the ellipticals were found to be less than 2 for both components, and therefore are not represented by de Vaucouleurs profiles. For J205050 and J225546, the components show $n_{inner}>n_{outer}$, while J020536 shows the opposite.  The axis ratio of the inner component is higher than the outer component in J020536 and J225546, while J205050 has nearly equal axis ratios for both components. For the single Sérsic fit, the Sérsic index is significantly higher ($n>2$) than either of the components of the double Sérsic fit. This higher Sérsic index reflects the attempt to fit both the steep and shallow light profiles dominating at different scales in the galaxy simultaneously. For a galaxy that has more than one component, this often results in an average value that does not represent either component accurately, but that rather provides a compromise whilst fitting the entire galaxy, leading to the impression of a highly compact object. Similarly, this affects the sizes of the galaxies, with a single Sérsic fit yielding an $R_e$ that falls between the $R_e$ values of the two components.

As is evident from the white-light images of the three ellipticals in Fig. \ref{fig:models}, there are a number of foreground stars or neighbouring galaxies within the field of view, which have the potential to affect the component models if not accounted for. For example, another object can be clearly observed close to J020536. \citet{lassen2021} concluded that this object was an irregular dwarf galaxy at $z=0.04025$ that is interacting with J020536, creating a plume of irregular structure between both objects. The asymmetric distortion in the surface brightness caused by the interaction makes the dwarf galaxy difficult to model simultaneously with the target. To avoid the fit from crashing or modelling the wrong object, we masked this entire region out with a polygon, in addition to masking out the stars. However, we note that despite our efforts to mask out the interacting object, the overlap between the galaxies implies that the light contamination from the dwarf galaxy in J020536 cannot be entirely eliminated, especially at the outskirts. Similarly, there is a small galaxy that appears close in projection to J225546. However, with a redshift of 0.22, it is clear that this object is neither a satellite galaxy accreting onto the target nor interacting with it. Rather, it is a background galaxy along the same line of sight as the main galaxy. However, since the projection of this object appears to lie very much within the outer regions of the target galaxy itself, it was decided to model this as an additional component. Since the `neighbouring' background galaxy is an extended object, discerning the point where its light contribution reaches negligible levels becomes challenging, making it difficult to mask. Choosing the size of the mask can have a significant effect in modelling the light from the main galaxy. In this instance, the mask could either be inadequate, allowing light from the outskirts to influence the fit to the target galaxy, or excessively large, resulting in substantial loss of light from the main galaxy. For these reasons, we elected to model the background galaxy, which helped us to subtract its light contribution, providing a more reliable fit to the galaxy of interest than masking it out would \citep{haeussler2007}.

The two-component models of the galaxies depicted in Fig. \ref{fig:models} show a good fit, barring the innermost regions. It must be noted that the seeing is indeed accounted for in these fits by convolving the flux datacubes with their PSF datacubes, but constructing an accurate PSF from IFU data presents a significant challenge, and it can often prove hard to model the central few pixels. The lower right panel depicts the fraction of light that has been over- or under-subtracted from the white-light image as a function of the galactocentric distance along the major axis. The majority of the residuals lie within $\pm1\%$ of the observed points for up to 20$''$ from the galactic centre. The higher scatter seen at larger distances from the centre are likely due to the asymmetric sky background and the effect of low S/N in the outskirts. The galaxy 1D profile was constructed by plotting the surface brightness from the centre of the observed galaxy to the outskirts using a simulated `slit' along the major axis, following the PA derived from \texttt{BUDDI} for the single Sérsic model. The model profiles were extracted with the same slit along the major axis, accounting for the PSF smearing at the centre in a consistent way for the inner and outer components and the combined model datacubes, which had all been convolved with the PSF in the initial steps of 2D fitting. The 1D and 2D surface brightness profiles can be considered a 1:1 correspondence, constructed from the white-light images collapsed along the wavelength dimension. While this serves to support the reliable quality of the fits to the surface brightness of the galaxies, it is important to note that the 1D profiles rarely capture mismatches with the surface brightness models (depending on the choice of sampling rate and  that are clearly visible spatially in the 2D profiles.

\subsection{Mean stellar populations}
\label{subsec:stellar_pops}

One of the main goals of this study is to investigate the physical significance of the two components that model the surface brightness profiles of elliptical galaxies, by examining their stellar properties. These dual-component models provide a more accurate representation than the single-component model that has been rooted in several decades of literature. The use of IFS allows us to study the stellar populations hosted by the two components, and identify any significant differences if they exist. This would ultimately help in correlating the stellar populations to physical mechanisms throughout the galaxies' lifetimes that have been driving the evolution of present-day classical ellipticals. 

\begin{figure}[h!]
    \centering
    \includegraphics[width=0.98\columnwidth]{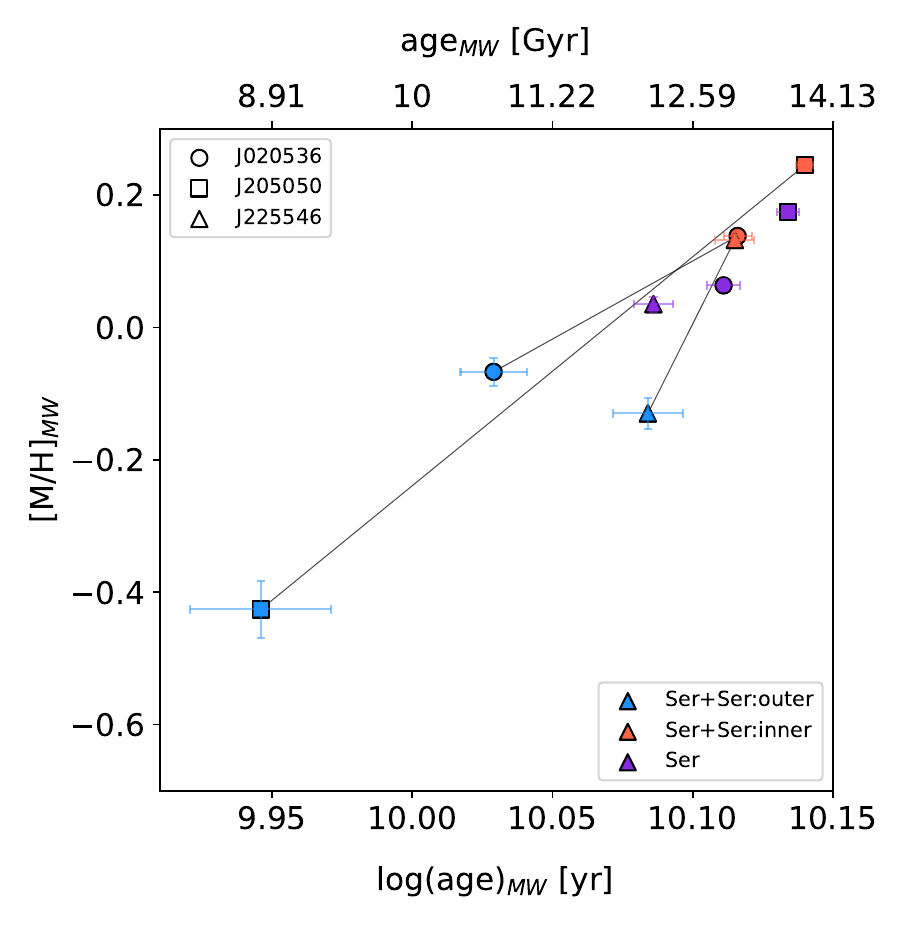}\\
    \includegraphics[width=0.98\columnwidth]{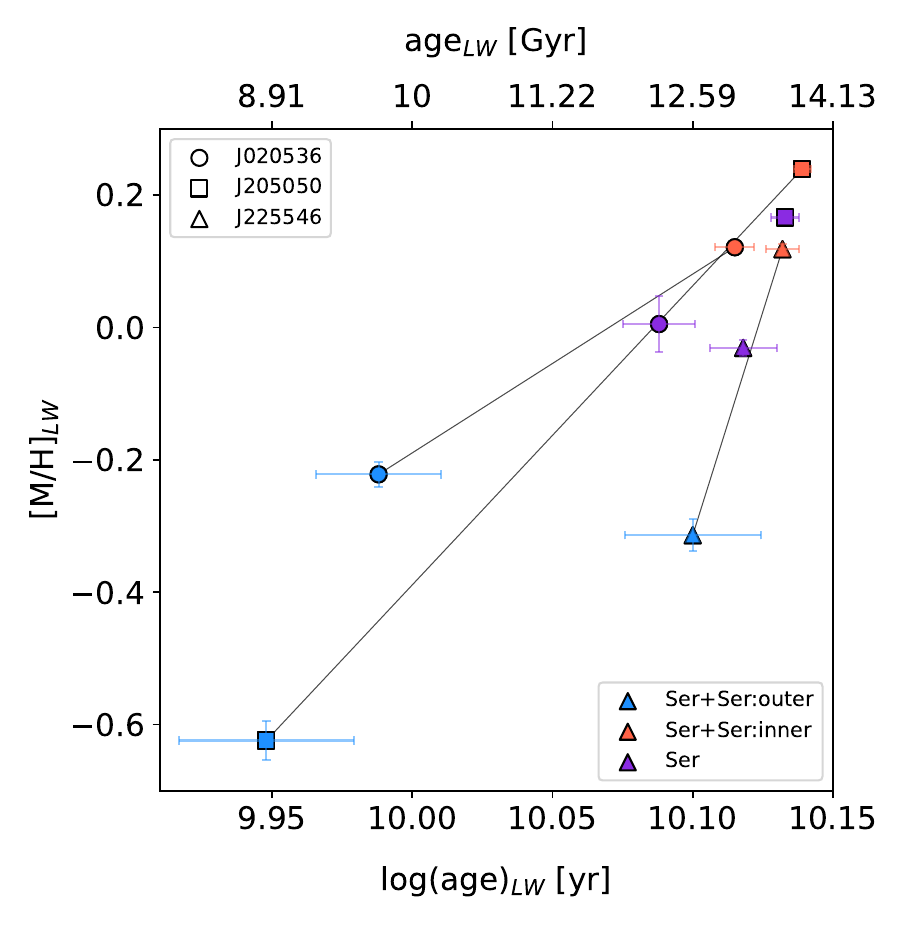}    
    \caption{Stellar populations derived from the single and two-component models using \texttt{pPXF}. Upper panel: Mass-weighted stellar populations showing the logarithmic stellar ages in years (lower $x$ axis) and their corresponding values in gigayears (upper $x$ axis), against stellar metallicities of the one-component and the two-component models. The three marker styles represent the three galaxies, while the colours represent the components - blue for the outer component of the double Sérsic model, red for the inner component, and purple for the single Sérsic component. The associated logarithmic error bars are also shown. Lower panel: Same as the upper panel, but with light-weighted stellar populations instead.}
    \label{fig:mw_lw_pops}
\end{figure}

Figure \ref{fig:mw_lw_pops} shows the stellar populations obtained from full spectral fitting with \texttt{pPXF} (see Sect. \ref{subsec:ppxf}). The mean mass-weighted stellar metallicities [M/H] and the logarithmic stellar ages, are depicted in the upper panel, while the mean light-weighted properties are shown in the lower panel. The circles, squares, and triangles represent the three elliptical galaxies in the sample (J020536, J205050, and J225546, respectively), and the red and blue colours mark the inner and outer components derived from \texttt{BUDDI} in the two-component (Sérsic + Sérsic) models. The inner and outer components of the same galaxy are connected by the black lines for better readability. For comparison, the stellar population properties from the standard single component (Sérsic) model are shown in purple.  Mass-weighted ages reflect the period when the majority of stellar mass was assembled and are less sensitive to the most luminous and younger stars; whereas light-weighted ages are influenced by recent star formation, which significantly impacts luminosity but contributes less to the overall mass \citep{Hopkins2018}.

The mean mass and light-weighted ages of the galaxies (single Sérsic) and their individual components (inner and outer components in the double Sérsic model) are all older than 9 Gyr, which typically places them in the category of long-term quiescent `classical ellipticals'. The difference between the ages of the populations present in the inner and outer components ranges roughly between 1 and 5 Gyr.  \citet{plauchu-frayn2012} define the star formation timescale $\Delta(t_*)$ as the difference between the mass-weighted and light-weighted stellar ages: $\Delta(t_*) = t_{MW} - t_{LW}$ Gyr. $\Delta(t_*)$ is high for objects that are continuously forming stars, while it is low for those that have ceased forming stars a long time ago. Following their definition, we compute $\Delta(t_*)$ for each component, as well as globally for each galaxy. Considering the ages of the galaxies as a whole, we find $\Delta(t_*)$ to be negligible, of the order of a few 100 Myr, well within their uncertainties.  This insignificant difference is typically expected for classical ellipticals when considered as a global property, implying that their stellar populations are homogeneous in age and would have formed and quenched at roughly similar early look-back times. The inner and outer components similarly show virtually imperceptible $\Delta(t_*)$ values. While we recognise that the mean light-weighted ages of J225546 appear to be slightly higher than the mass-weighted ages outside their error bounds, it is important to reiterate that one of the major caveats of the fossil-record method is that the stellar ages are indistinguishable for extremely old stellar populations. Furthermore, since the chosen isochrones of the SSP models used in the spectral fitting include ages up to $\sim14$ Gyr, \texttt{pPXF} can estimate mass fractions at ages older than the Hubble time, depending on the assumed cosmology. Due to these limitations, we exercise caution in directly interpreting the individual absolute stellar ages and the star formation timescale. From the mean stellar ages, it is only clear that the $\Delta(t_*)$ values are low, and therefore suggest populations that are similar in age. However, the quantity $\Delta(t_*)$ is more related to intrinsic star formation and quenching in the galaxy, and less sensitive to dry mergers where stellar material is simply accreted without a subsequent star forming episode, unless the accreted galaxy itself is young. Therefore, the low values of the outer component do not necessarily confirm that the population was formed at the same time throughout the component, but could be attributed to stellar accretion from satellite galaxies that are approximately as old as the main galaxy.  

The mean mass-weighted stellar metallicities of the galaxies indicate metal-rich stellar populations, with super-solar metal abundances ([M/H]$_{MW}$>0). For the inner components, the metallicities are quite high compared to the global values for the galaxies. In contrast, the outer components signify metal-poor stellar populations, with sub-solar metal abundances ([M/H]$_{MW}$<0). Compared to the outer components, the inner components are significantly more metal-enhanced by factors of $0.2-0.7$ dex. Similarly, the light-weighted metallicities decrease inside-out, with the outer components showing less metal-enhanced stellar populations than their corresponding inner components by factors of $0.3-0.9$ dex. The relative differences between the mass-weighted and the light-weighted metallicities $\Delta[M/H]_*$ are quite small, between 0 and 0.2 dex, with the highest differences exhibited by the outer components. Our stellar populations analysis indicates a lack of recent star formation and younger stars, resulting in the mass and light-weighted maps appearing quite similar, with no concentrations of recent star formation. Any differences between the mass and light-weighted quantities are likely due to variations in metallicities rather than ages. These results align with the two-phase formation scenario where the inner component has likely assembled most of its stellar mass through in situ bulk star formation when the interstellar medium was already enriched with metals, either through dissipative collapse of a gas cloud or through major mergers between gas-rich discs. This component would then accrete its stellar mass later in dry mergers of relatively metal-poor galaxies, such as dwarf galaxies, or through star formation resulting from relatively pristine gas inflow that dilutes the metal-enhanced interstellar medium.

\begin{figure}
    \centering
    \includegraphics[width=0.9\columnwidth]{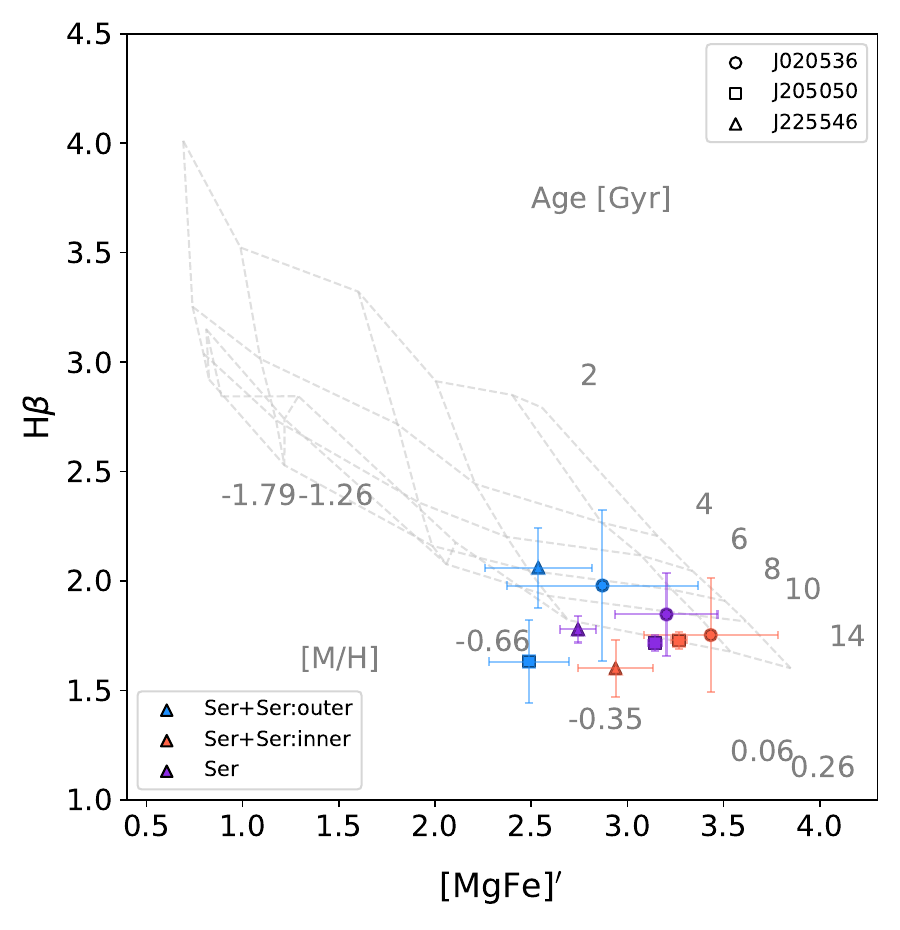}
    \caption{Absorption line strengths and associated error bars of the age and metallicity indicators: H$\beta$, Mg$b$, and Fe, measured with \texttt{PyLick}. As in Fig. \ref{fig:mw_lw_pops}, the three marker styles indicate the different galaxies, while the red and blue colours indicate the inner and outer component of each galaxy, respectively. The points are overlaid on the model SSP grid in grey.} 
    \label{fig:lick_ind}
\end{figure}

The consistency check for the stellar populations measurements using Lick indices are shown in Fig. \ref{fig:lick_ind}. The overlaid grid in grey represents the model SSP measurements from \citet{vazdekis2010}, and are conventionally used to convert the index measurements into physical stellar population properties. We have not done the latter conversion step, since most of the points lie outside the grid, instead only observing the qualitative implications of these results in tandem with the full-spectral fitting analysis. Nevertheless, the grid provides a guide to a ballpark estimate of the stellar age and metallicity. A cursory inspection of the points in the H$\beta$ - [MgFe]$'$ plane and their locations within the model SSP grid reveals that the inner components are typically older and more metal-rich than the outer components. However, the error bars are quite significant to conclusively determine this systematic difference, or if the inner components within each galaxy are older and more metal-enhanced compared to their respective outer components. The approximate stellar populations from this analysis align with the conclusions of those derived from the more extensive full-spectral fitting method. 

At this point, it is important to consider the possible caveats of employing the limited optical wavelength range of MUSE in deriving stellar populations. Given that the spectral range of the instrument extends only to the H$\beta$ absorption line in the blue end, this could lead to the implication that the lack of higher-order Balmer lines makes the analysis less sensitive to very recent star formation episodes at timescales $t<100$ Myr. We note that while the H$\delta$ and H$\gamma$ would be indispensable in breaking the age-metallicity degeneracy when only the absorption line strengths are used in deriving stellar populations, this effect is mitigated in full-spectral fitting where multiple spectral features sensitive to both age and metallicity are modelled simultaneously alongside the stellar continuum. For this reason, and the fact that several galaxy components fall outside the model SSP grid in the H$\beta$ - [MgFe]$'$ plane, we conducted the rest of the stellar population analysis using the full-spectral fitting method.

\begin{table*}[]
\centering
\caption{Mean mass and light-weighted stellar populations properties listing the stellar metallicities, ages, and formation times, $\tau_{50}$ and $\tau_{90}$.}
\begin{tabular}{@{}cccccc@{}}
\hline \hline
\multicolumn{6}{c}{\textbf{Mean mass-weighted stellar populations}}                                                                                           \\ \midrule
\textbf{Galaxy}                   & \textbf{Comp} & \textbf{{[}M/H{]}} & \textbf{Age (Gyr)} & \textbf{$\bm{\tau_{50}}$ (Gyr)} & \textbf{$\bm{\tau_{90}}$ (Gyr)} \\ \midrule
\multirow{3}{*}{\textbf{J020536}} & single             & 0.063 $\pm$ 0.009             &  12.91 $\pm$ 0.180              & 13.42 $\pm$ 0.266                      & 11.90 $\pm$ 1.138                    \\
                                  & inner              & 0.138 $\pm$ 0.006             & 13.06 $\pm$ 0.164             & 13.50 $\pm$ 0.315                     & 12.11 $\pm$ 0.894                   \\
                                  & outer              & -0.067 $\pm$ 0.021             & 10.69 $\pm$ 0.297             & 11.41 $\pm$ 0.902                     & 8.28 $\pm$ 0.827                     \\ \midrule
\multirow{3}{*}{\textbf{J205050}} & single             & 0.174 $\pm$ 0.004             & 13.61 $\pm$ 0.116             & 13.96 $\pm$ 0.063                     & 13.17 $\pm$ 0.408                    \\
                                  & inner              & 0.245 $\pm$ 0.004             & 13.80 $\pm$ 0.099             & 14.00 $\pm$ 0.025                     & 13.55 $\pm$ 0.387                   \\
                                  & outer              & -0.426 $\pm$ 0.043            & 8.83 $\pm$ 0.553             & 9.26 $\pm$ 1.247                    & 6.71 $\pm$ 0.905                    \\ \midrule
\multirow{3}{*}{\textbf{J225546}} & single             & 0.035 $\pm$ 0.011              & 12.19 $\pm$ 0.200             & 12.81 $\pm$ 0.489                     & 10.57 $\pm$ 1.125                    \\
                                  & inner              & 0.132 $\pm$ 0.007             & 13.03 $\pm$ 0.222             & 13.50 $\pm$ 0.250                     & 12.09 $\pm$ 0.826                   \\
                                  & outer              & -0.130 $\pm$ 0.024             & 12.13 $\pm$ 0.347             & 12.83 $\pm$ 0.565                     & 10.41 $\pm$ 1.080                     \\
\multicolumn{6}{c}{}                                                                                                                                     \\
\hline \hline
\multicolumn{6}{c}{\textbf{Mean light-weighted stellar populations}}                                                                                          \\ \midrule
\textbf{Galaxy}                   & \textbf{Comp} & \textbf{{[}M/H{]}} & \textbf{Age (Gyr)} & \textbf{$\bm{\tau_{50}}$ (Gyr)}   & \textbf{$\bm{\tau_{90}}$ (Gyr)} \\ \midrule
\multirow{3}{*}{\textbf{J020536}} & single             & 0.005 $\pm$ 0.042            & 12.25 $\pm$ 0.349             & 13.44 $\pm$ 0.355                     & 11.86 $\pm$ 1.610                   \\
                                  & inner              & 0.121 $\pm$ 0.006            & 13.03 $\pm$ 0.207              & 13.47 $\pm$ 0.345                     & 12.06 $\pm$ 0.860                  \\
                                  & outer              & -0.222 $\pm$ 0.019            & 9.73 $\pm$ 0.492              & 10.73 $\pm$ 0.751                     & 6.77 $\pm$ 1.394                    \\ \midrule
\multirow{3}{*}{\textbf{J205050}} & single             & 0.166 $\pm$ 0.003             & 13.58 $\pm$ 0.143            & 13.94 $\pm$ 0.045                     & 13.15 $\pm$ 0.453                   \\
                                  & inner              & 0.239 $\pm$ 0.004             & 13.77 $\pm$ 0.112             & 14.00 $\pm$ 0.015                     & 13.55 $\pm$ 0.378                   \\
                                  & outer              & -0.624 $\pm$ 0.029             & 8.87 $\pm$ 0.617             & 8.97 $\pm$ 1.093                     & 7.25 $\pm$ 1.395                    \\ \midrule
\multirow{3}{*}{\textbf{J225546}} & single             & -0.031 $\pm$ 0.012            & 13.12 $\pm$ 0.347             & 13.57 $\pm$ 0.226                     & 12.18 $\pm$ 1.113                    \\
                                  & inner              & 0.118 $\pm$ 0.008             & 13.55 $\pm$ 0.200               & 13.93 $\pm$ 0.137                     & 13.10 $\pm$ 0.805                    \\
                                  & outer              & -0.314 $\pm$ 0.024            & 12.59 $\pm$ 0.642              & 13.55 $\pm$ 0.557                     & 10.51 $\pm$ 1.584                   \\  \midrule \\ 
\end{tabular}

\label{tab:all_stellar_pops}
\end{table*}

\subsection{Comparison with spatially resolved 2D stellar population maps}
\label{subsec:voronoi_SP}

\begin{figure*}
    
  \centering
  \begin{tabular}{cccc}
    \includegraphics[trim=1.5cm 0.8cm 3.8cm 0cm, clip, width=0.265\textwidth]{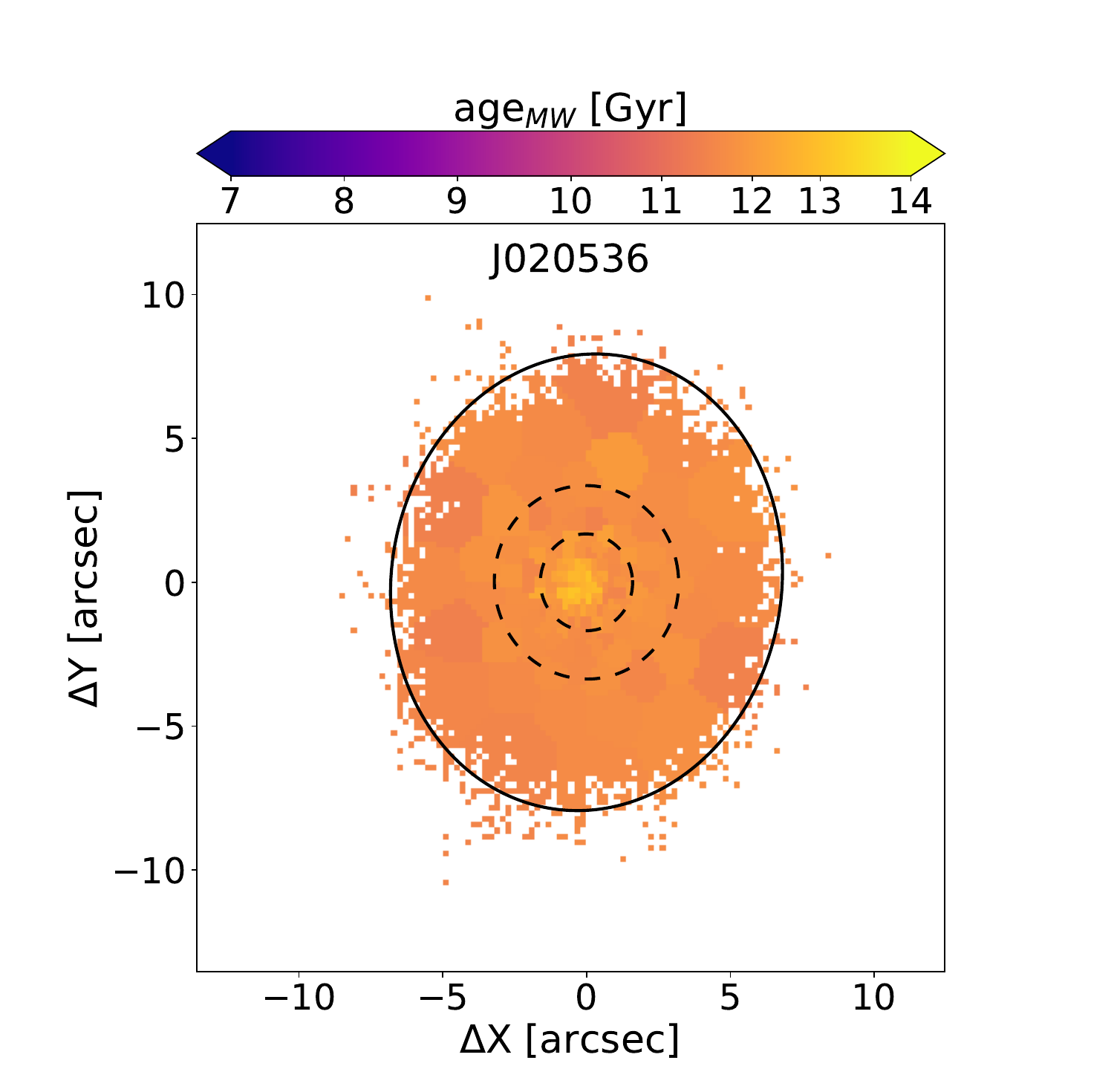} & 
    \includegraphics[trim=1.5cm 0.8cm 3.8cm 0cm, clip, width=0.265\textwidth]{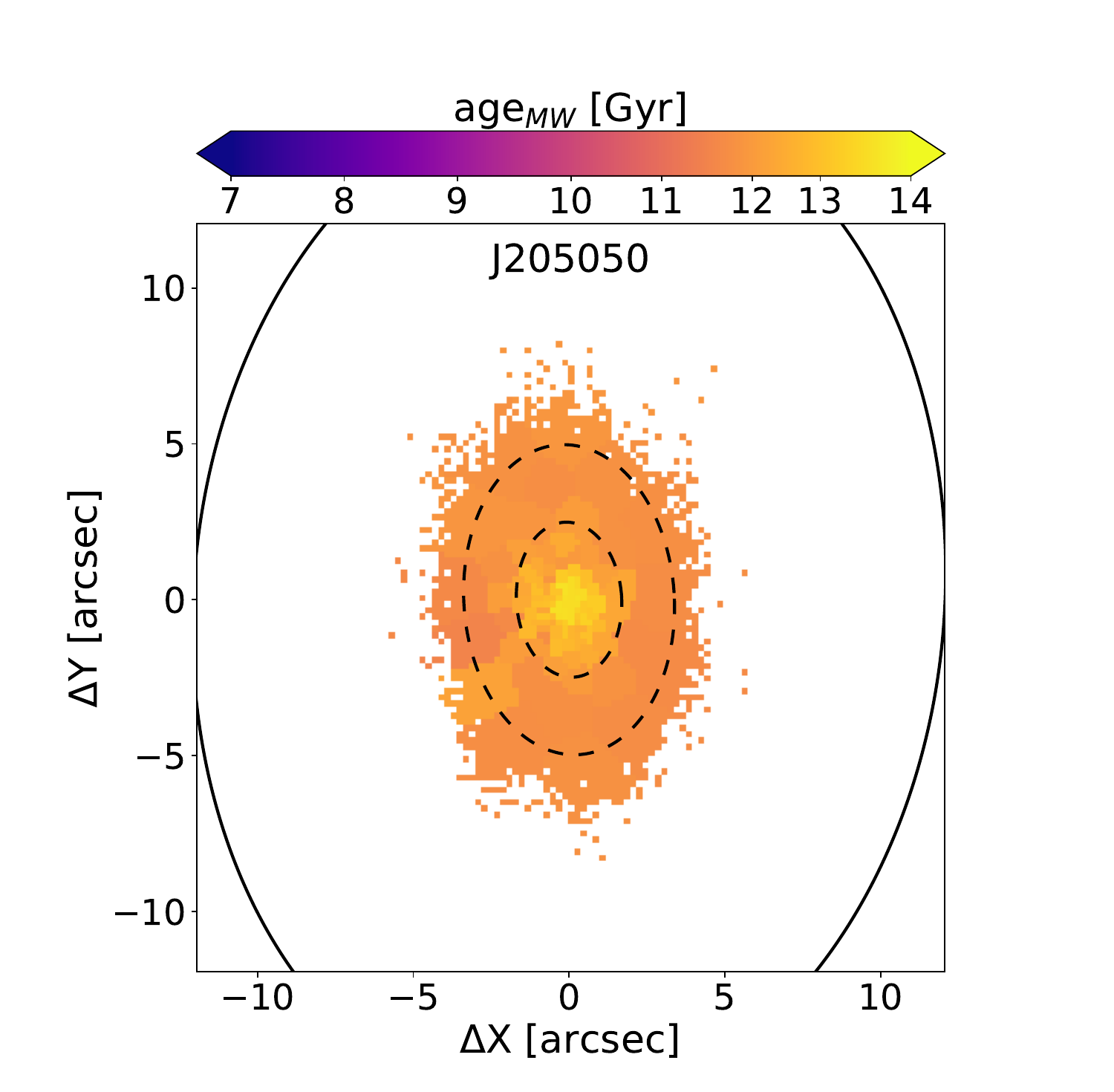} & 
    \includegraphics[trim=1.5cm 0.8cm 3.8cm 0cm, clip, width=0.265\textwidth]{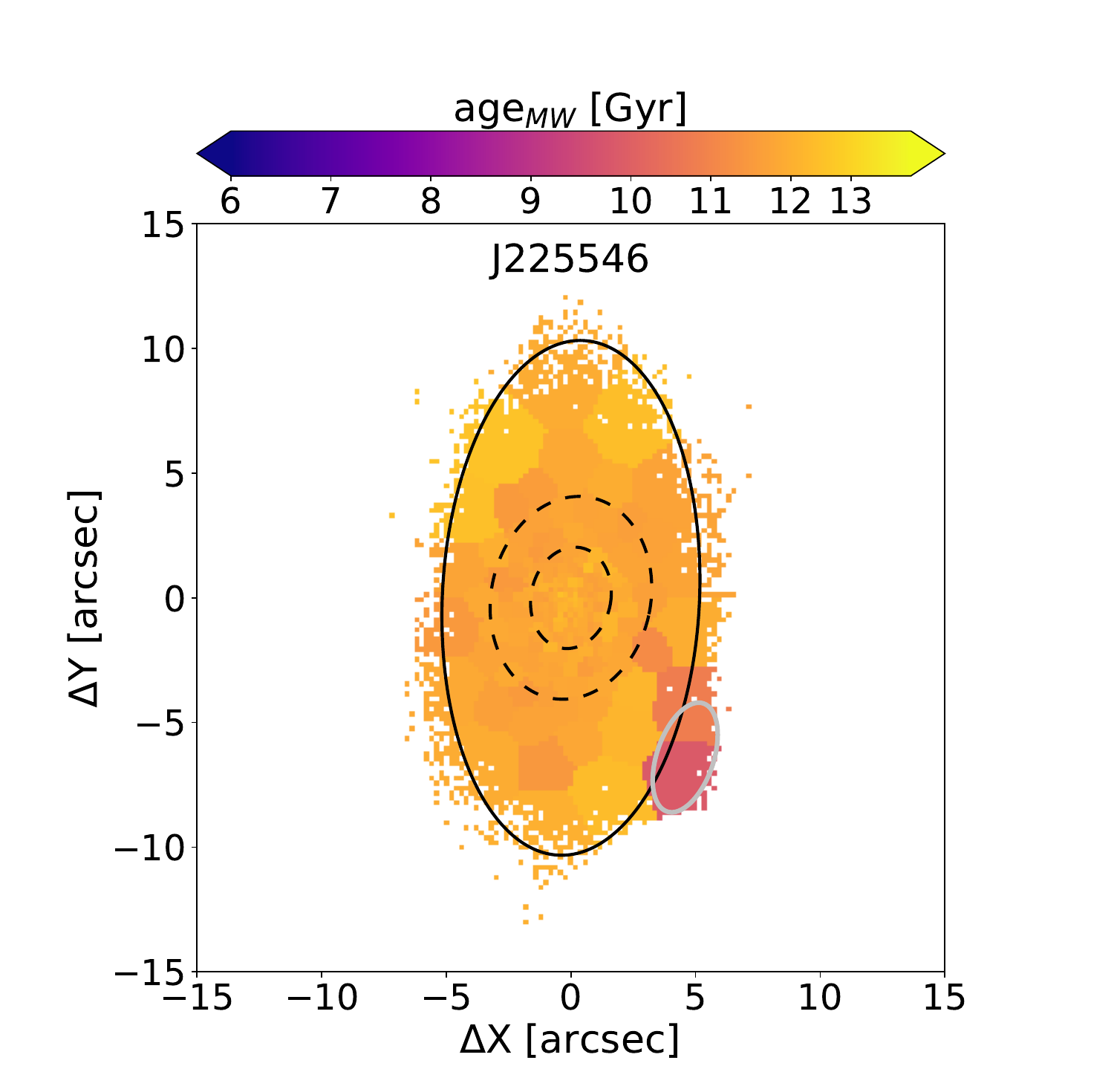} \\ 
    \includegraphics[trim=1.5cm 0.8cm 3.8cm 0cm, clip, width=0.265\textwidth]{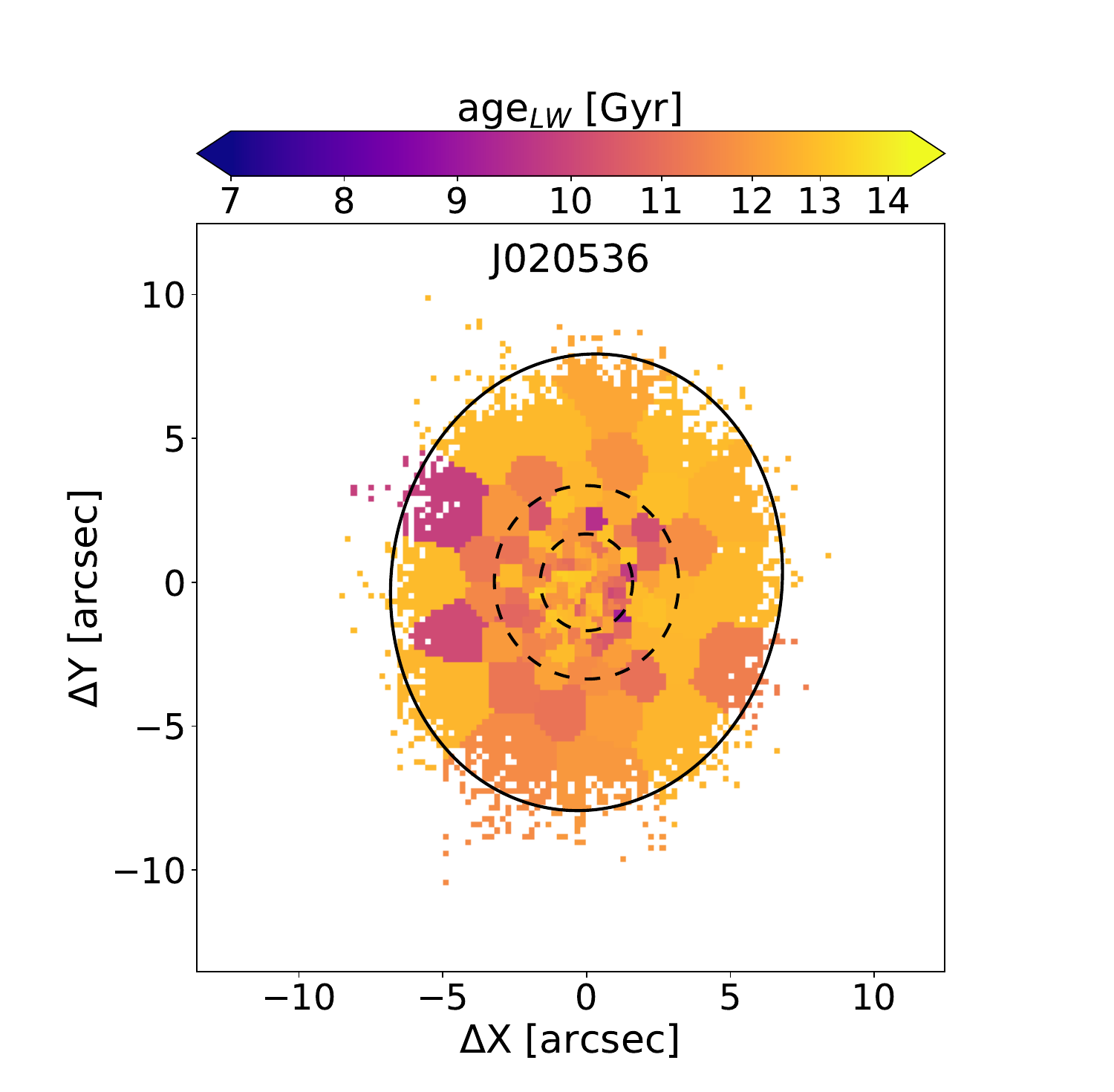} &
    \includegraphics[trim=1.5cm 0.8cm 3.8cm 0cm, clip, width=0.265\textwidth]{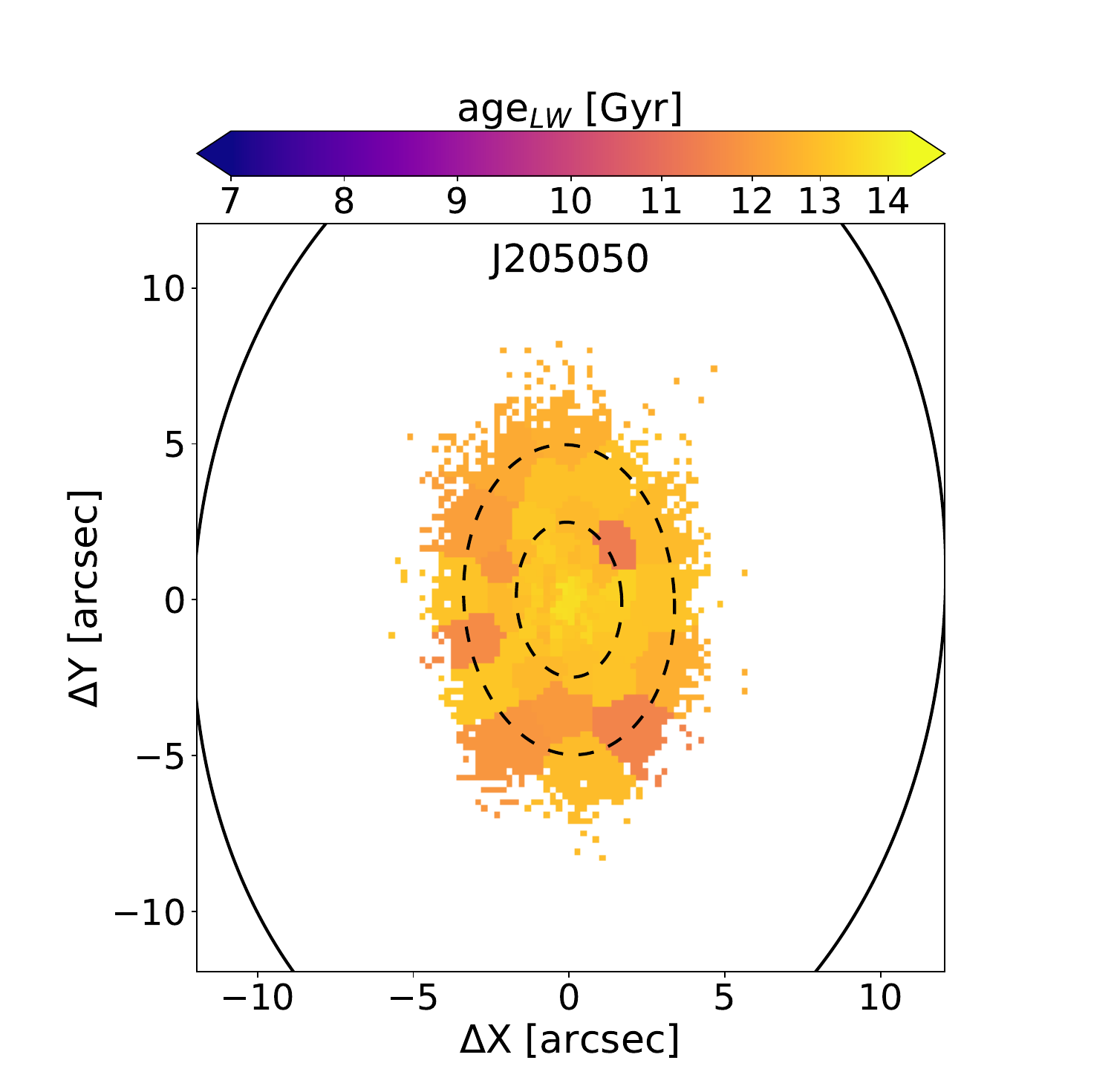} & 
    \includegraphics[trim=1.5cm 0.8cm 3.8cm 0cm, clip, width=0.265\textwidth]{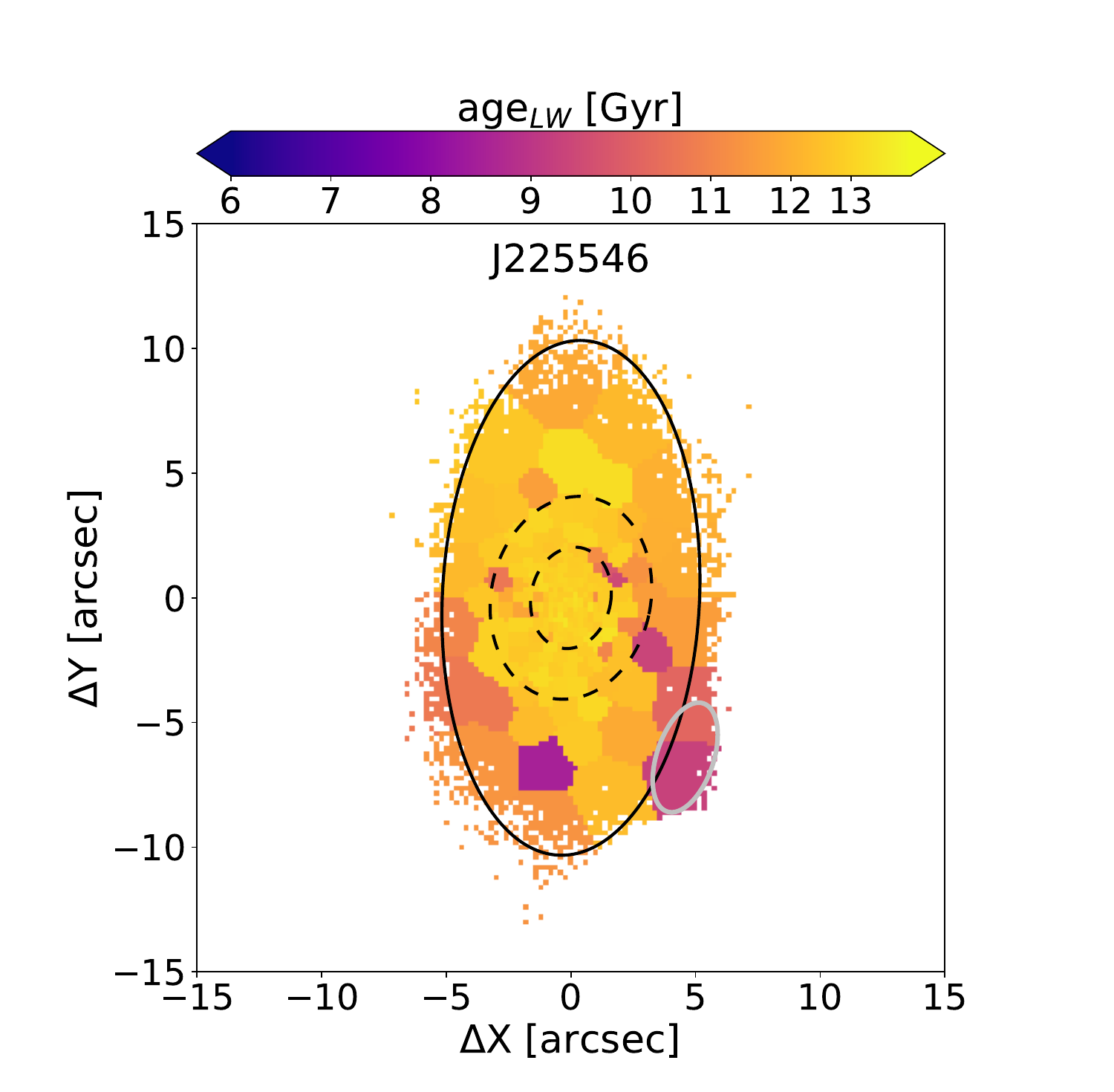} \\ 
    \includegraphics[trim=1.5cm 0.8cm 3.8cm 0cm, clip, width=0.265\textwidth]{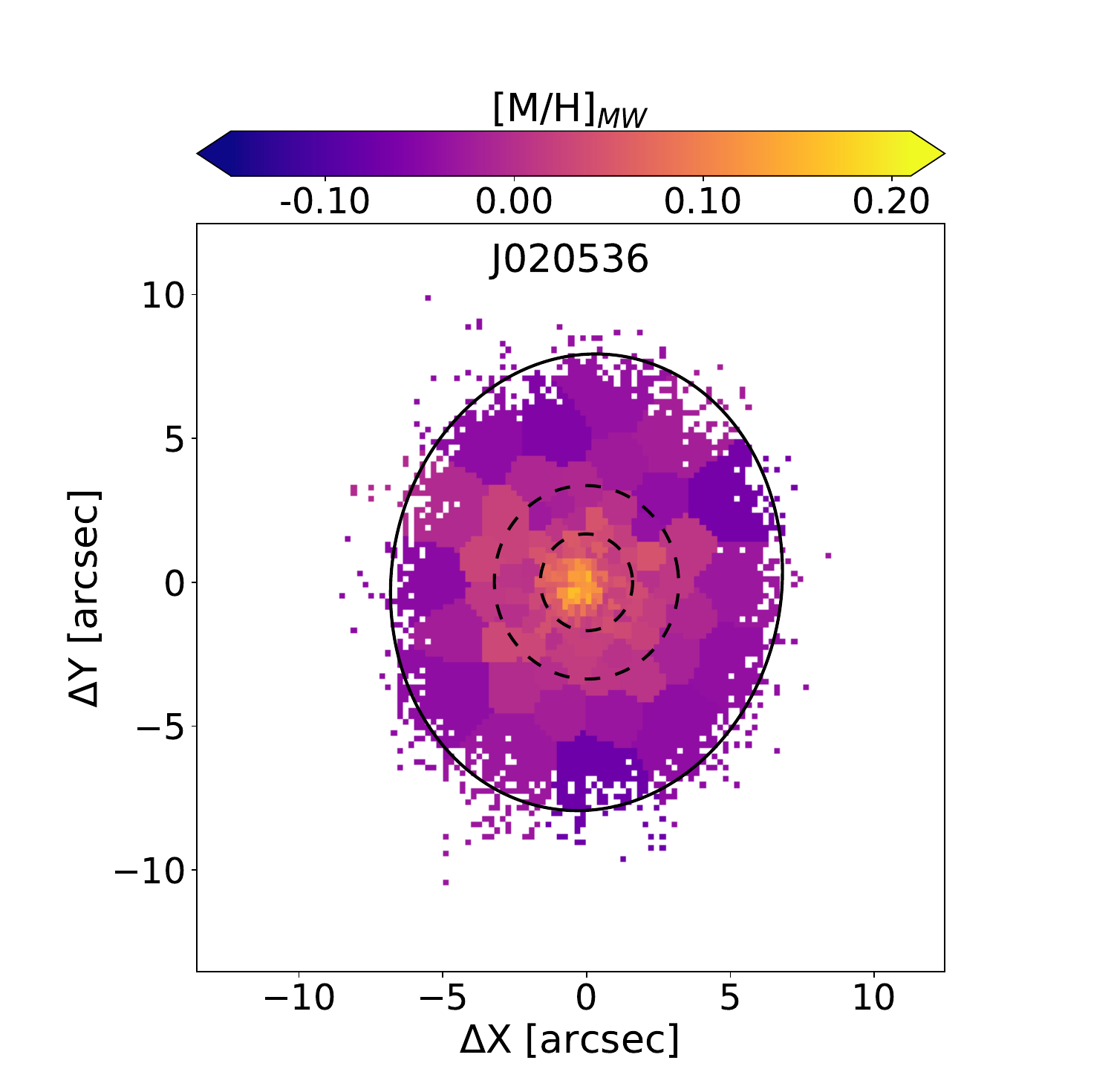} & 
    \includegraphics[trim=1.5cm 0.8cm 3.8cm 0cm, clip, width=0.265\textwidth]{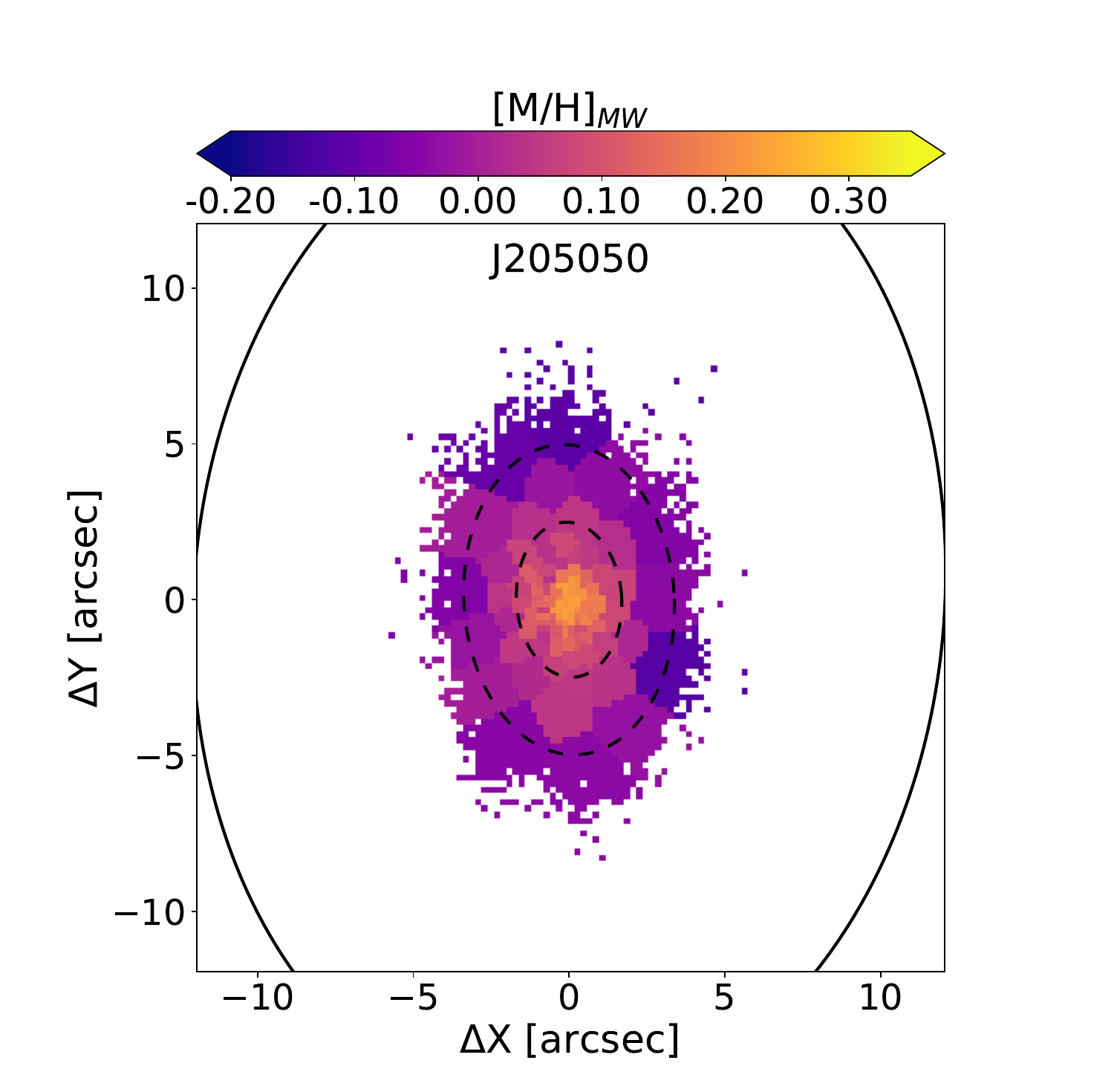} &
    \includegraphics[trim=1.5cm 0.8cm 3.8cm 0cm, clip, width=0.265\textwidth]{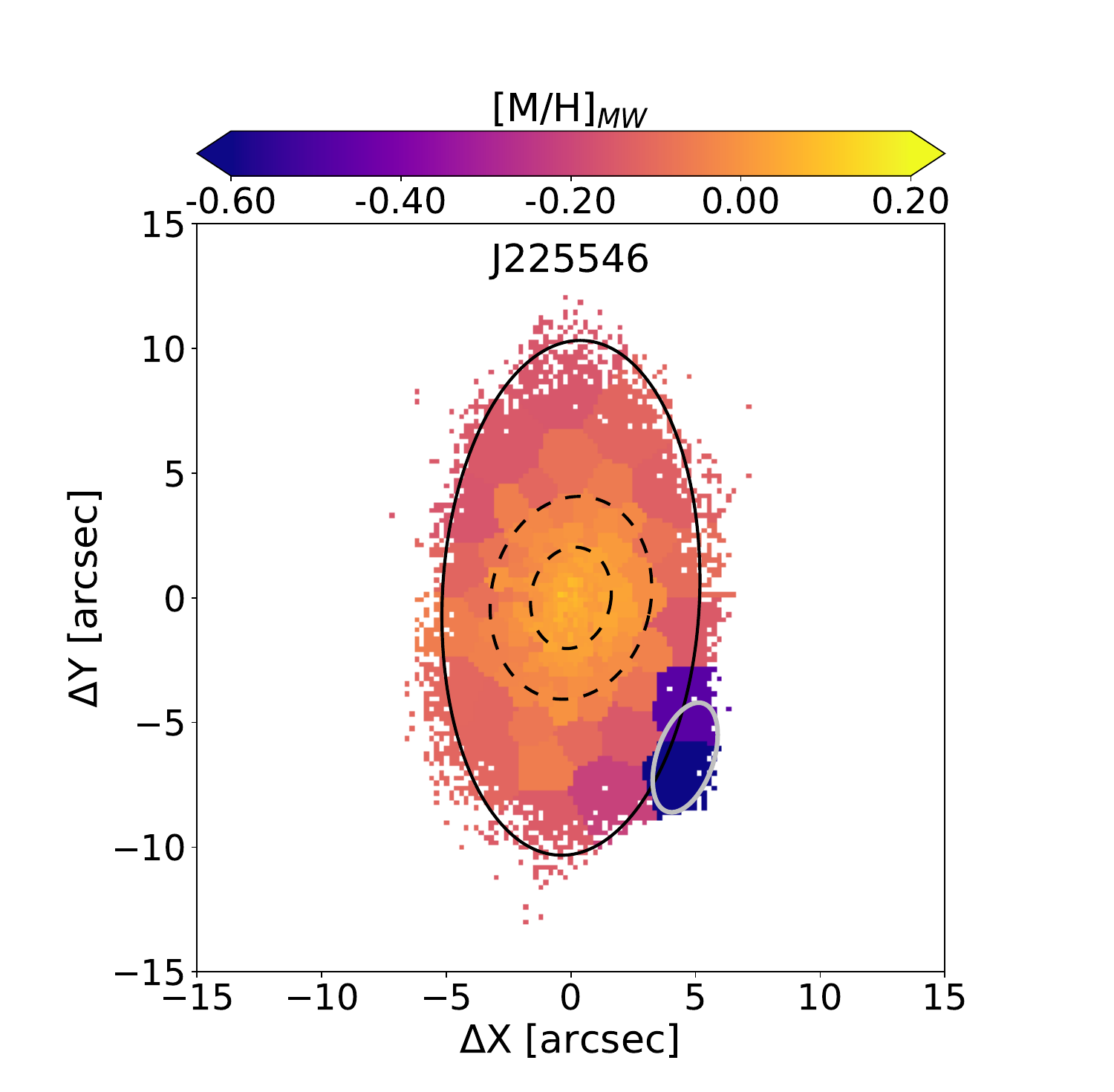} \\ 
    \includegraphics[trim=1.5cm 0.8cm 3.8cm 0cm, clip, width=0.265\textwidth]{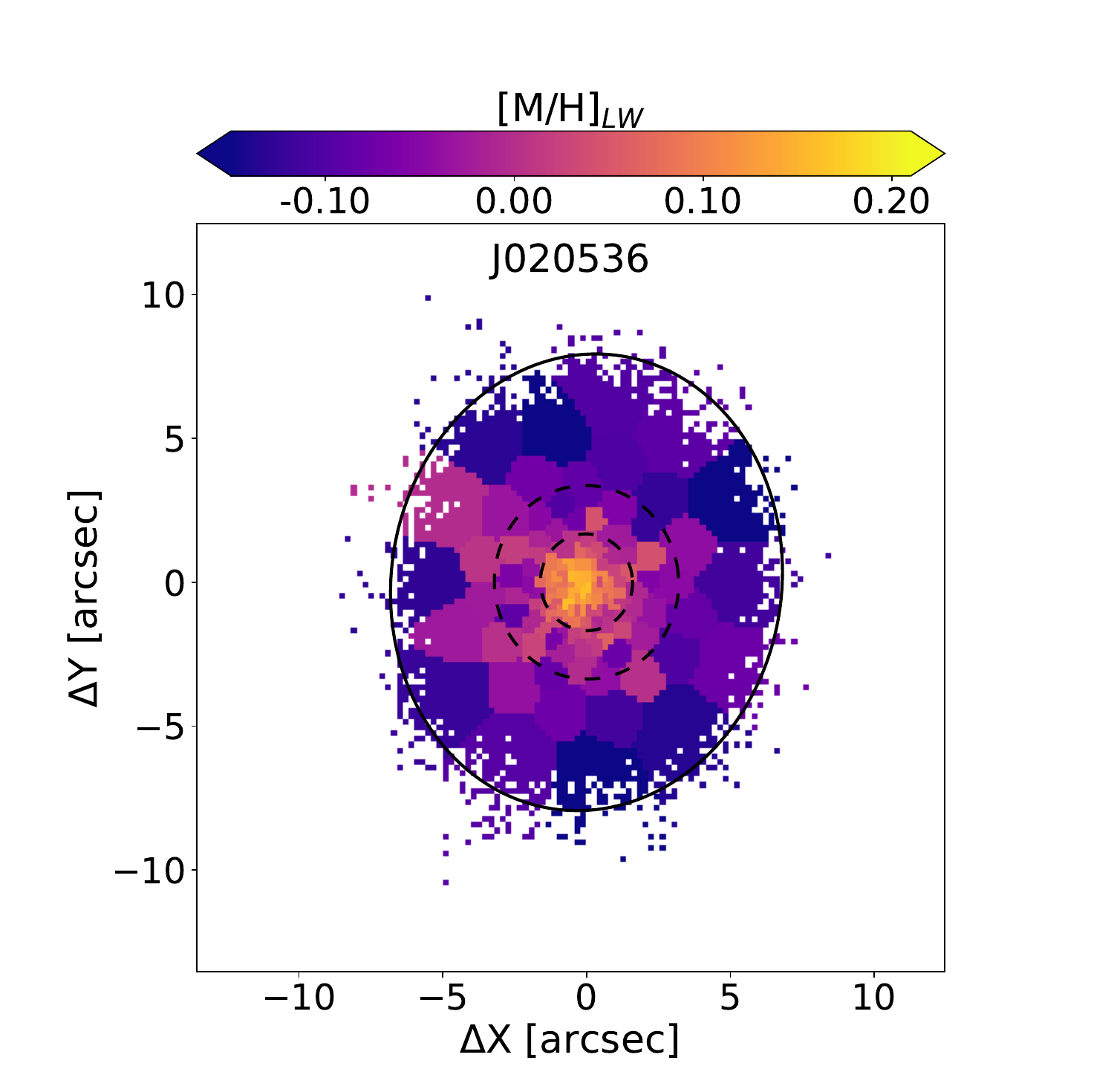} & 
    \includegraphics[trim=1.5cm 0.8cm 3.8cm 0cm, clip, width=0.265\textwidth]{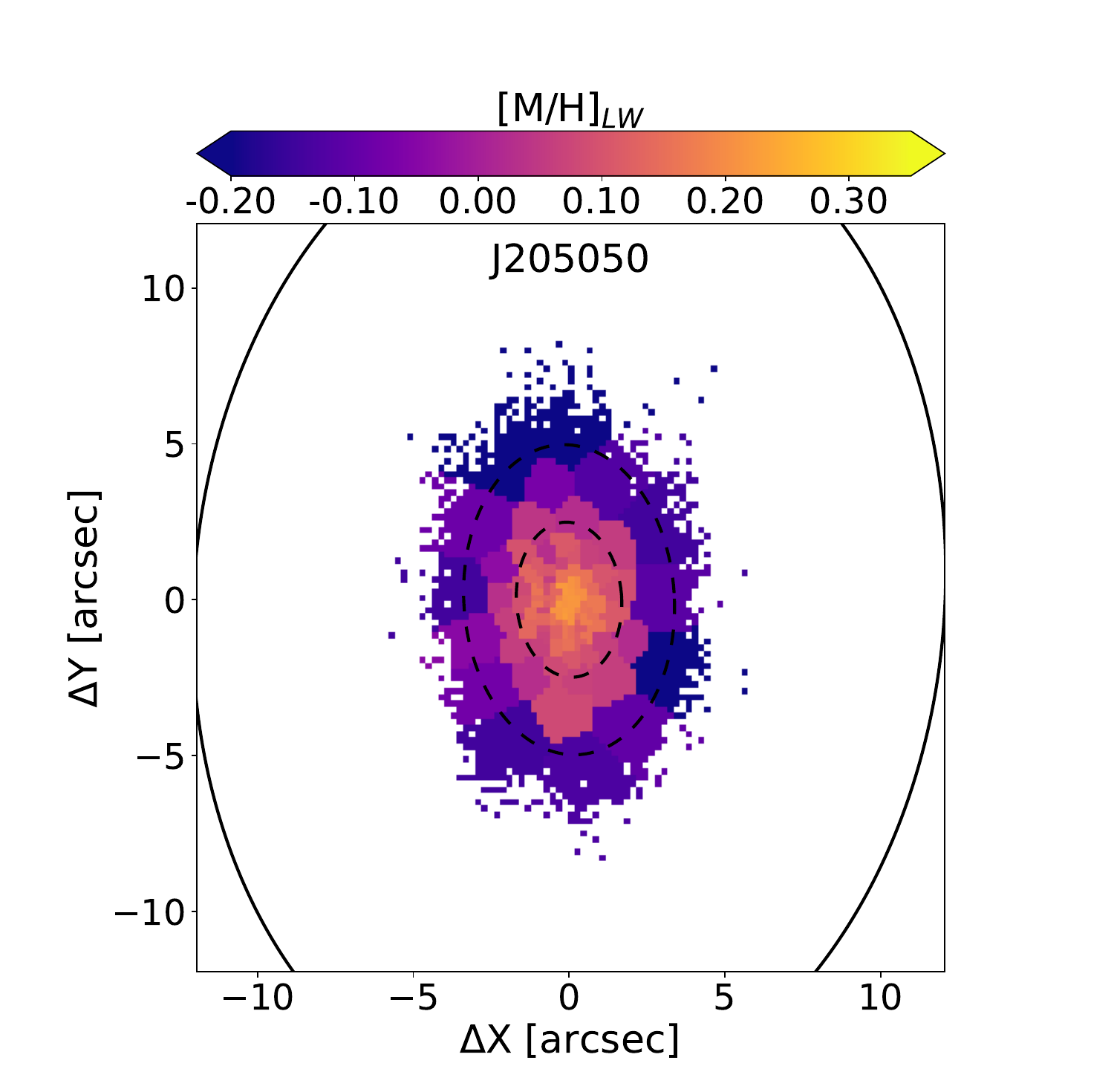} & 
    \includegraphics[trim=1.5cm 0.8cm 3.8cm 0cm, clip, width=0.265\textwidth]{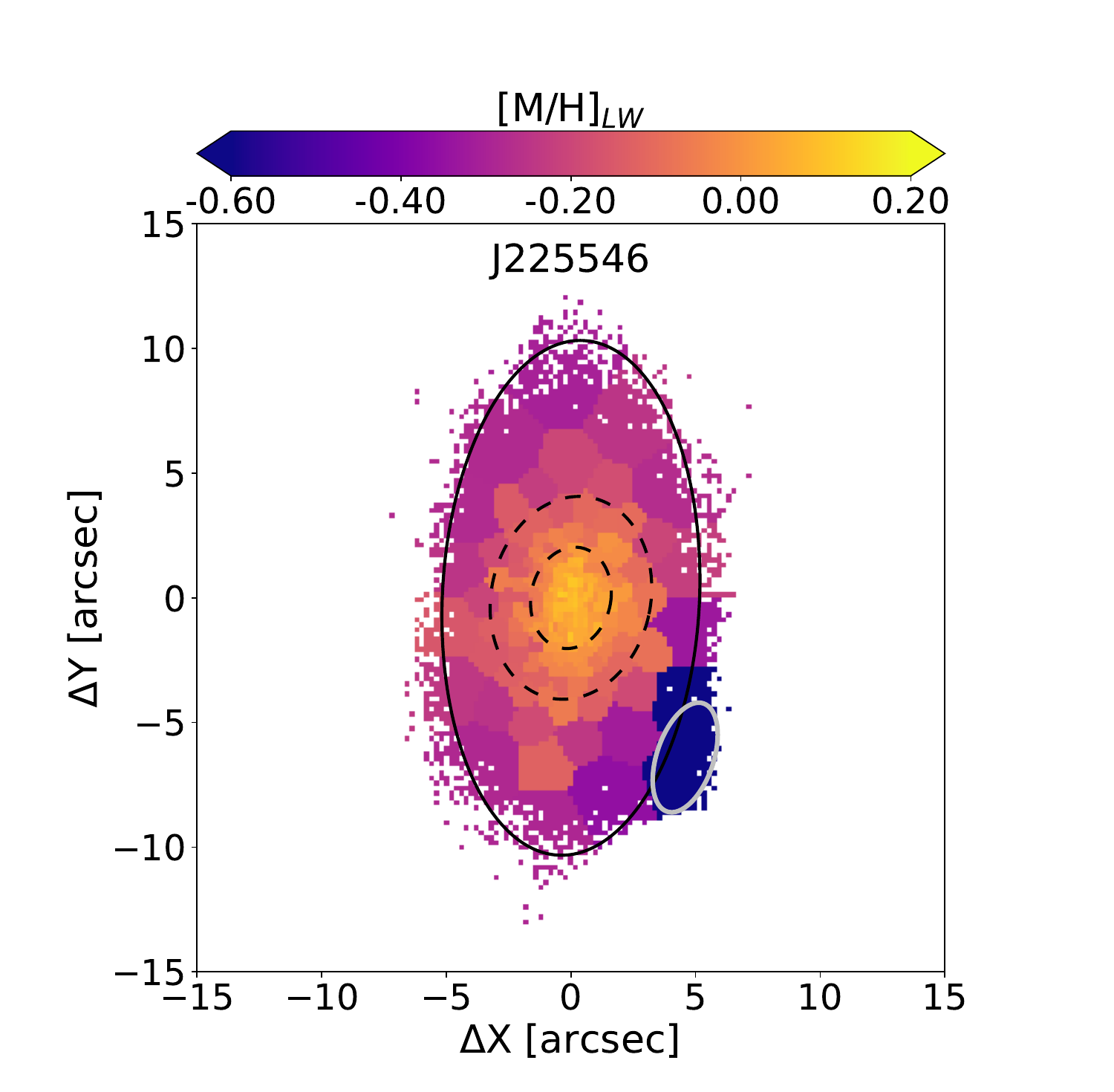} \\
  \end{tabular}
  
  \caption{Voronoi-binned stellar populations for galaxies J020536, J205050, and J225546 from left to right. From top to bottom, the rows indicate the logarithmic mass-weighted stellar ages, the light-weighted ages, the mass-weighted metallicities, and the light-weighted metallicities. The colour bars for the ages have been displayed in gigayears for better clarity, although the maps themselves are shown with the logarithmic ages (causing the uneven spacing of the linear ages in gigayears for the colour bar). The 1$R_e$ and 2$R_e$ contours from \texttt{BUDDI} of the inner component are marked as dashed black ellipses, and only the 1$R_e$ contour of the outer component is shown as solid black ellipses, since the 2$R_e$ contours are beyond the extent of the mapped regions of the galaxy. For object J225546, an additional contour is shown in grey, which marks the 1$R_e$ ellipse of the adjacent galaxy modelled together with the target.}
  \label{fig:voronoi_SP}
\end{figure*}

We also conducted a complementary stellar population analysis to compare the global properties described in Sect. \ref{subsec:stellar_pops}. For this, the Voronoi binning technique was used to bin the IFU datacubes based on their S/N, and the spectra in each bin were summed together to retrieve a single spectrum with adequate S/N (we refer the reader to Sect. \ref{subsec:voronoi} for details). The binned spectra were then fitted with \texttt{pPXF} using the same method detailed in Sect. \ref{subsec:ppxf}. This technique allows us to create a 2D map of the stellar population properties and observe qualitatively any radial trends present in the galaxy. However, as is stated in Sect. \ref{subsec:voronoi}, these do not serve as a direct comparison to the results obtained with \texttt{BUDDI} due to the significant chance of mixing between different structural components. Nevertheless, these 2D maps offer an alternative way of comparing our results, albeit in a more generic way, and resemble more traditional analyses in the literature. The columns in Fig. \ref{fig:voronoi_SP} represent the three elliptical galaxies, while each row shows a single stellar population property: mass-weighted age, light-weighted age, mass-weighted metallicity, and light-weighted metallicity, in that order. The properties are mapped based on the colour bar above each plot, with the youngest ages and lowest metallicities in indigo, and the oldest ages and the highest metallicities in yellow. The black contours mark the 1$R_e$ and 2$R_e$ ellipses for the inner (dashed) and outer (solid) components modelled in the two-component fit to each galaxy.

For galaxy J020536, the mass-weighted stellar age map shows a mildly negative gradient from the central region, corresponding to $1R_e$ of the inner component derived from \texttt{BUDDI}, to the outskirts, corresponding to $1R_e$ of the outer component. We find the negative mass-weighted age gradients to be composed of a dominant old ($t\gtrsim12$ Gyr) stellar population in the inner component, and a slightly younger but homogeneous population ($t\sim11$ Gyr) on the outskirts. The light-weighted stellar age map, however, does not appear to show any significant trends, similar to the findings of \citet{lassen2021}, where the spatially resolved light-weighted mean ages were derived through SPS at the spaxel resolution. The mass-weighted and light-weighted stellar metallicity maps, on the other hand, show a very clear gradient from the centre to the outskirts. The core hosts the most metal-rich population in the galaxy, which becomes less enhanced moving radially outwards towards the galactic outskirts. This trend is reflected in the inner component derived from \texttt{BUDDI}, with a negative gradient progressing from 1$R_e$ to 2$R_e$. This trend continues even moving towards 1$R_e$ of the outer component, which hosts the least metal-enhanced populations in the galaxy. While both the mass-weighted and the light-weighted metallicity maps systematically show the same trend, the light-weighted map also appears to show less metal-enhanced populations compared to the mass-weighted map.

Likewise, J205050 appears to host a central core that is older than the rest of the galaxy, which also lies within 1$R_e$. Beyond the central region, however, the mass-weighted stellar age map is mostly homogeneous. From the light-weighted age map, it is difficult to observe any gradient, there appear to be both younger and older stellar populations across the outer component of the galaxy, but the oldest populations are still clustered in the centre. Both the mass-weighted and light-weighted metallicity maps effectively mirror those presented in galaxy J020536. The central core shows relatively higher metallicities, with a decline moving radially outwards, again reflected in the modelled components moving from 1$R_e$ to 2$R_e$ of the inner component. The trends exhibited in the light-weighted metallicity map are again only different in the sense that they are systematically less metal-enhanced than in the mass-weighted map.  

The mass-weighted age map of J225546 does not indicate any substantial systematic trends, exhibiting both older stellar populations ($t\gtrsim12$ Gyr) and slightly younger populations with differences hardly reaching 1 Gyr across different regions of the galaxy. The light-weighted age map similarly shows a diverse range of stellar ages across the galaxy. In both cases, the region south-west of the galaxy shows an anomalous population compared to the rest of the galaxy outskirts or the outer component. This area corresponds to the location of the galaxy at $z=0.22$, where the Voronoi bins are dominated by its strong emission rather than J225546, causing this region to appear younger. The metallicity maps are consistent with the previously described galaxies, with the central region dominated by metal-rich stars, and a negative gradient with lower metallicity stars being hosted in the outskirts. Similarly, the most metal-enhanced regions in both the mass-weighted and light-weighted maps are hosted within 2$R_e$ of the modelled inner component; although the negative gradient is steeper in the light-weighted map than it is in the mass-weighted map moving from 1$R_e$ to 2$R_e$. We also note that the neighbouring galaxy hosts relatively metal-poor stellar populations; however, analysing this galaxy is beyond the scope of this work.

\subsection{Temporal mass and luminosity contributions from inner and outer components}
\label{subsec: mass_light_weights}

\begin{figure*} 
    
  \centering
  \begin{tabular}{cccc}
    \includegraphics[width=0.3\textwidth]{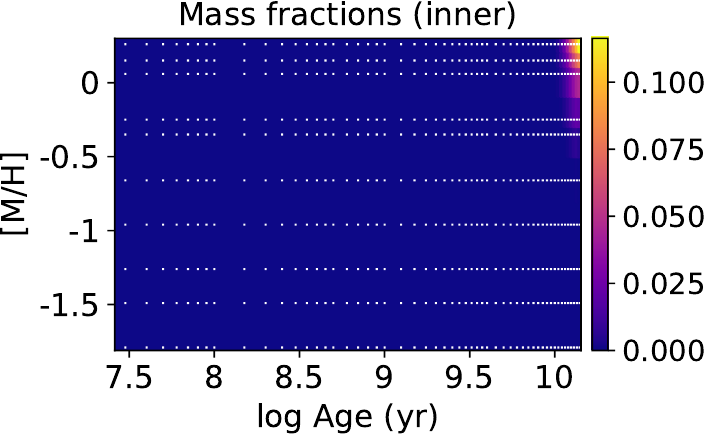} & 
    \includegraphics[width=0.3\textwidth]{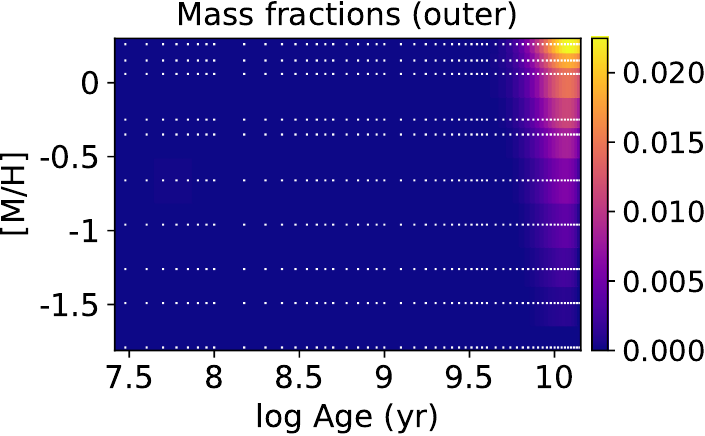} & 
    \includegraphics[width=0.3\textwidth]{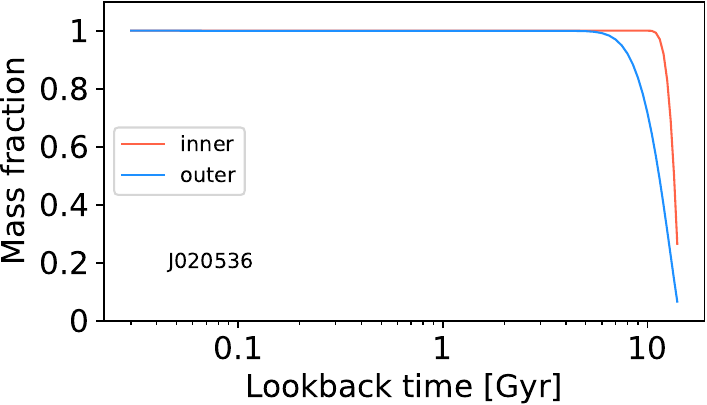} \\ \\
    \includegraphics[width=0.3\textwidth]{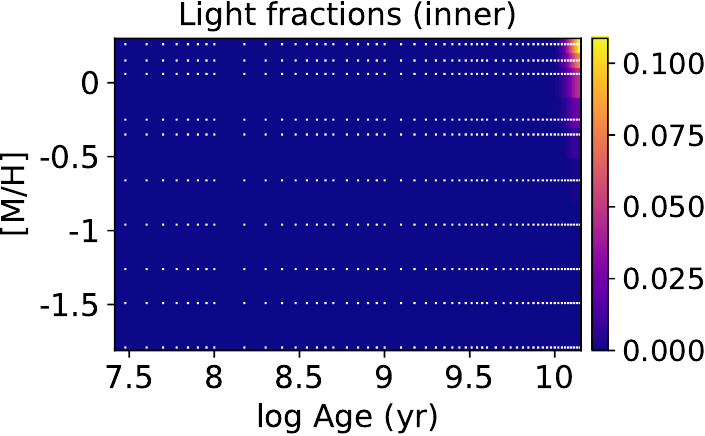} &
    \includegraphics[width=0.3\textwidth]{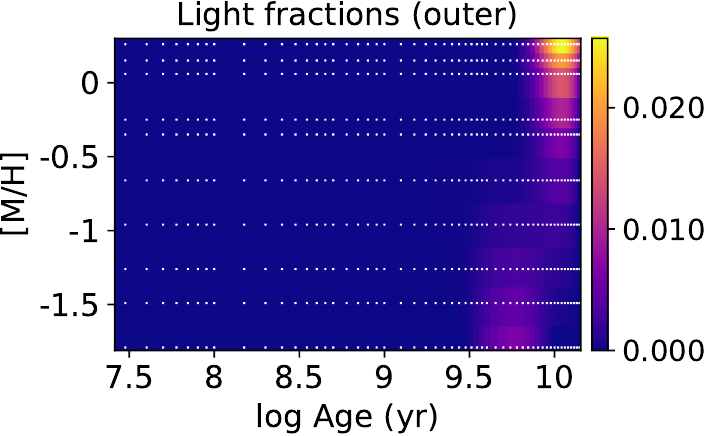} & 
    \includegraphics[width=0.3\textwidth]{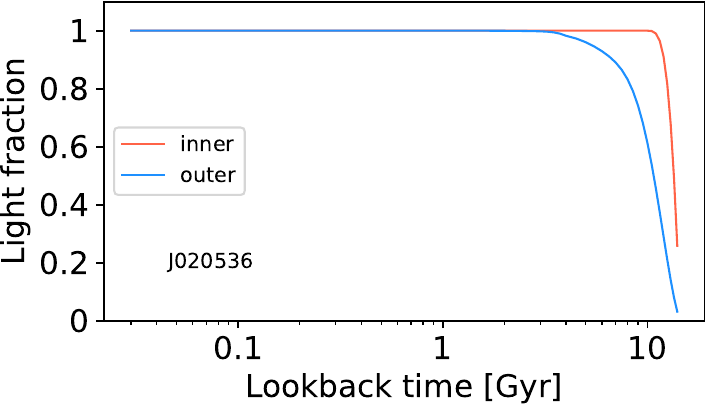} \\ 
  \end{tabular}
  \caption{Stellar populations in metallicity-age grids and their subsequent assembly histories for galaxy J020536.}
  \label{fig:j020536_assembly}
\end{figure*}

\begin{figure*} 
    
  \centering
  \begin{tabular}{cccc}
    \includegraphics[width=0.3\textwidth]{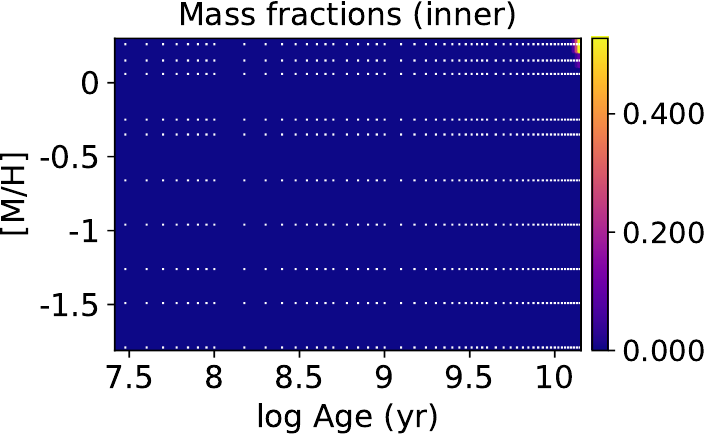} & 
    \includegraphics[width=0.3\textwidth]{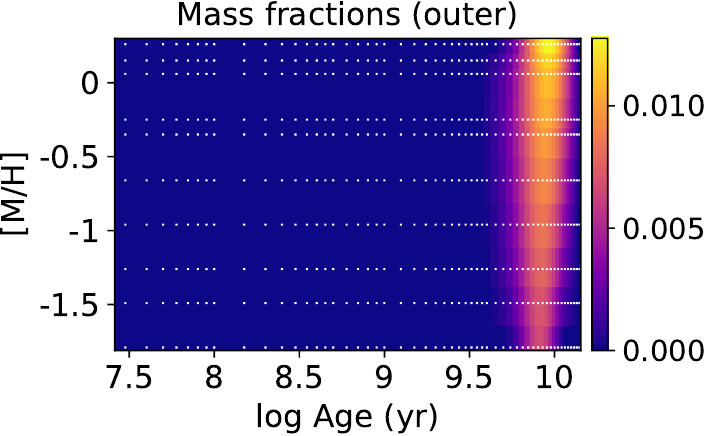} & 
    \includegraphics[width=0.3\textwidth]{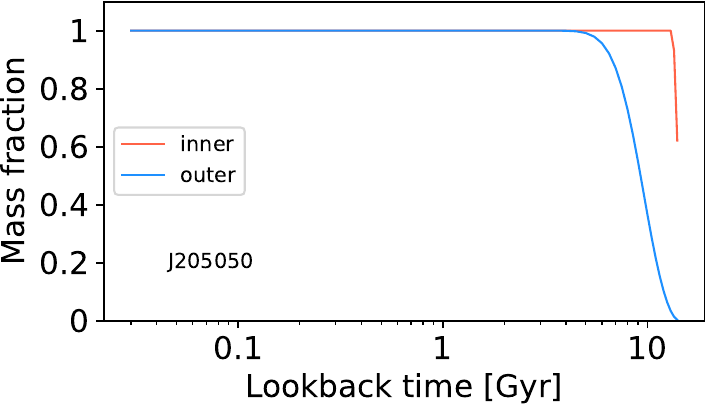} \\ \\
    \includegraphics[width=0.3\textwidth]{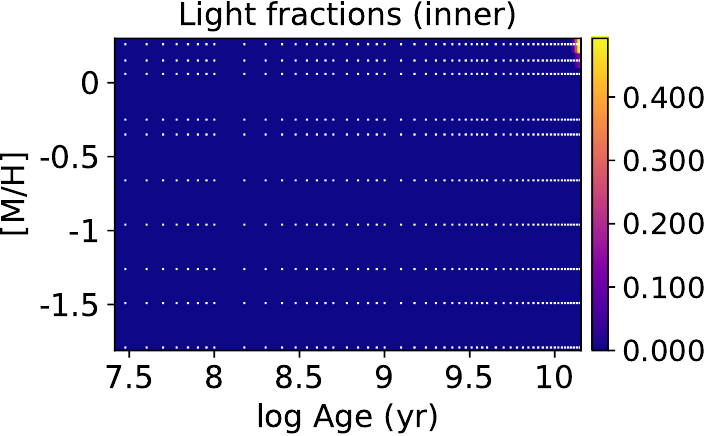} &
    \includegraphics[width=0.3\textwidth]{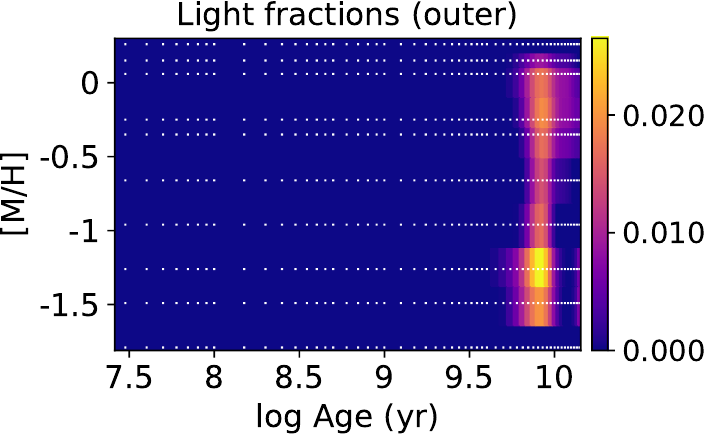} & 
    \includegraphics[width=0.3\textwidth]{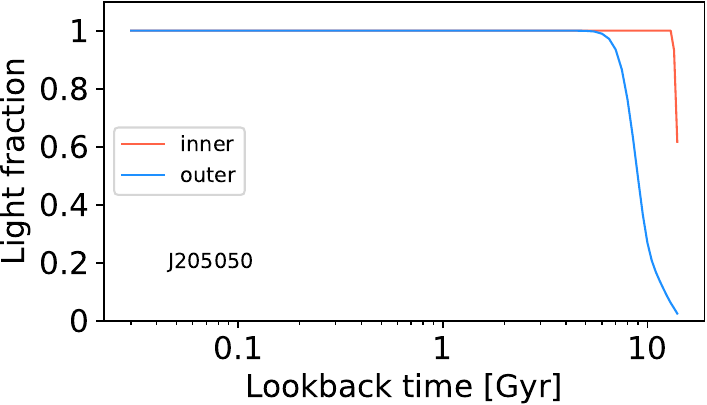} \\ 
  \end{tabular}
  \caption{Stellar populations in metallicity-age grids and their subsequent assembly histories for galaxy J205050.}
  \label{fig:j205050_assembly}
\end{figure*}

\begin{figure*} 
    
  \centering
  \begin{tabular}{cccc}
    \includegraphics[width=0.3\textwidth]{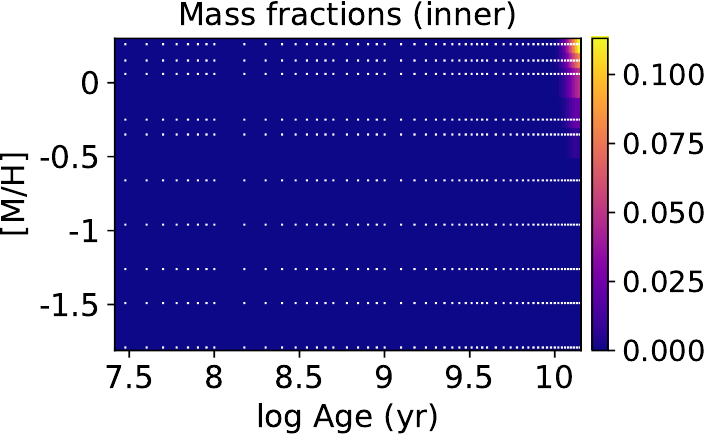} & 
    \includegraphics[width=0.3\textwidth]{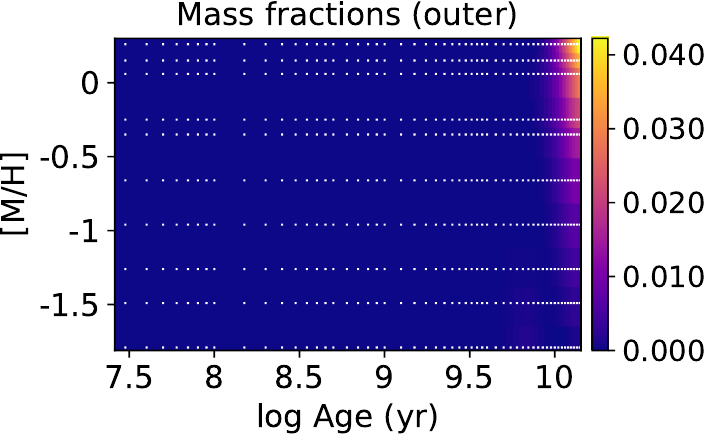} & 
    \includegraphics[width=0.3\textwidth]{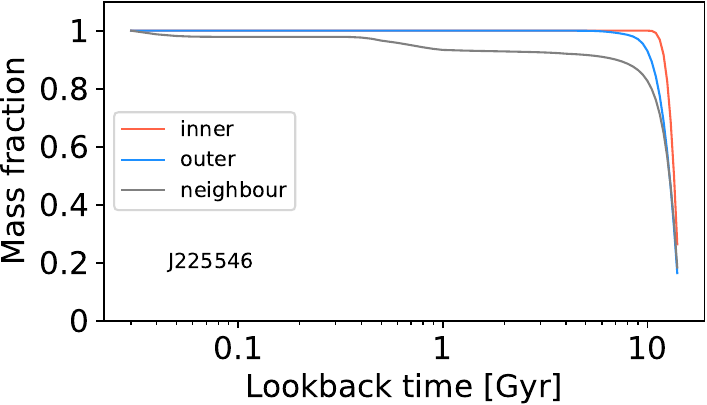} \\ \\
    \includegraphics[width=0.3\textwidth]{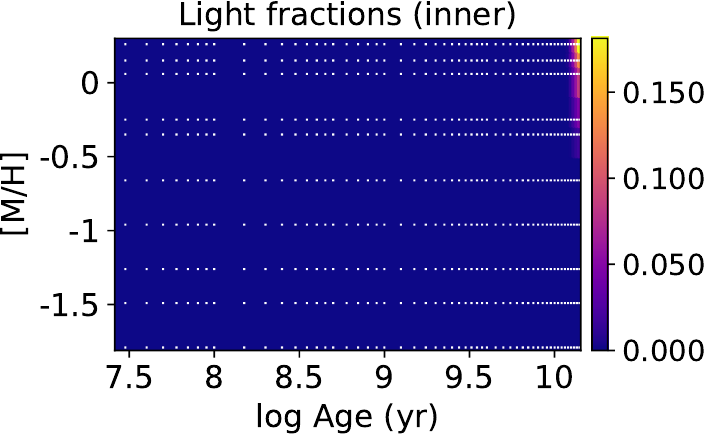} &
    \includegraphics[width=0.3\textwidth]{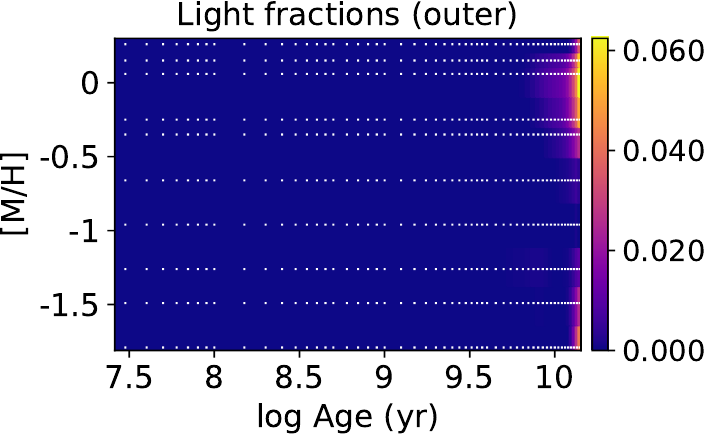} & 
    \includegraphics[width=0.3\textwidth]{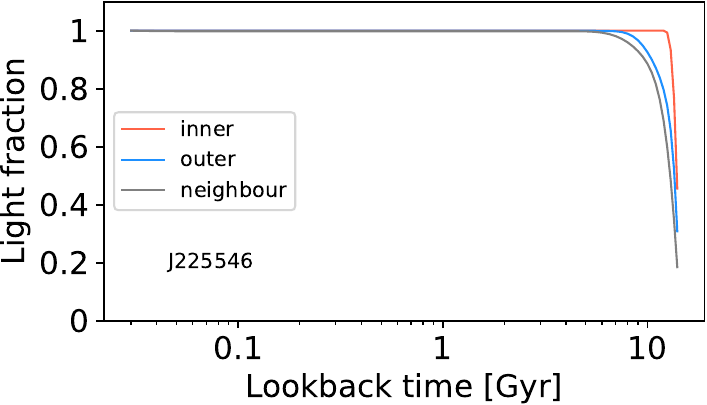} \\ 
  \end{tabular}
  \caption{Stellar populations in metallicity-age grids and their SFHs for galaxy J225546.}
  \label{fig:j225546_assembly}
\end{figure*}

The full spectral fitting process with \texttt{pPXF} returns a set of weights representing the relative contribution of each input SSP template to the optimised best-fit solution. From this, the 2D SFH of the galaxy components can be plotted on an age-[M/H] grid created from the stellar age and metallicity steps of the input stellar templates (see first and second columns in Figs. \ref{fig:j020536_assembly}, \ref{fig:j205050_assembly}, and \ref{fig:j225546_assembly}). This proxy of SFH portrays the chemical evolution of the components through time, indicated by episodes of star formation. The 1D SFHs (see third column in Figs. \ref{fig:j020536_assembly}, \ref{fig:j205050_assembly}, and \ref{fig:j225546_assembly}) can then be constructed by cumulatively summing up the normalised weights over all the metallicities at each stellar age step in the grid, such that the total fraction is 1. In this study, since we analyse both mass-weighted and light-weighted stellar populations (Sect. \ref{subsec:stellar_pops}), our reconstructed SFHs highlight the mass fraction and light fraction formed at different look-back times. The time evolution of the cumulative mass fraction provides insights into the stellar mass assembly history of the galaxy, such as whether it formed stars rapidly in the early universe or experienced prolonged star formation. The cumulative light fraction, on the other hand, highlights any recent star formation prominently since massive, young stars dominate the stellar luminosity of the galaxy and therefore have a higher impact on the light distribution when compared to the cumulative mass fraction. Furthermore, the light-weighted age distributions trace the most active phase of star formation in the galaxy.

\begin{table}[]
\centering
\caption{Mass fractions and light fractions estimated by \texttt{pPXF}.}
\begin{tabular}{@{}ccccc@{}}
\hline \hline
\multicolumn{5}{c}{\textbf{Mass fractions (\%)}}                                                                                                    \\ \midrule
\textbf{Galaxy}                   & \textbf{Comp}   & \textbf{$\bm{t_{old}}$} & \textbf{$\bm{t_{inter}}$} & \textbf{$\bm{t_{young}}$} \\ \midrule
\multirow{3}{*}{\textbf{J020536}} & single & 99.96                      & 0                         & 0.04                            \\
                                  & inner  & 100                      & 0                         & 0                         \\
                                  & outer  & 88.34                      & 11.58                         & 0.08                         \\ \midrule
\multirow{3}{*}{\textbf{J205050}} & single & 100                      & 0                            & 0                         \\
                                  & inner  & 100                      & 0                            & 0                         \\
                                  & outer  & 64.13                      & 35.83                       & 0                            \\ \midrule
\multirow{3}{*}{\textbf{J225546}} & single & 99.58                      & 0.42                         & 0                         \\
                                  & inner  & 100                      & 0                         & 0                         \\
                                  & outer  & 98.01                      & 1.99                         & 1.50                         \\ 
\multicolumn{5}{c}{}                                                                                                                           \\

\hline \hline
\multicolumn{5}{c}{\textbf{Light fractions (\%)}}                                                                                                   \\ \midrule
\textbf{Galaxy}                   & \textbf{Comp}   & \textbf{$\bm{t_{old}}$} & \textbf{$\bm{t_{inter}}$} & \textbf{$\bm{{t_{young}}}$} \\ \midrule
\multirow{3}{*}{\textbf{J020536}} & single & 98.53                      & 0                        & 1.47                         \\
                                  & inner  & 100                      & 0                         & 0                         \\
                                  & outer  & 79.18                      & 19.07                        & 0.34                         \\ \midrule
\multirow{3}{*}{\textbf{J205050}} & single & 100                      & 0                            & 0                         \\
                                  & inner  & 100                      & 0                            & 0                         \\
                                  & outer  & 63.19                      & 36.81                        & 0                            \\ \midrule
\multirow{3}{*}{\textbf{J225546}} & single & 100                      & 0                         & 0                       \\
                                  & inner  & 100                      & 0                         & 0                         \\
                                  & outer  & 98.04                      & 1.96                         & 0                         \\ \midrule \\
\end{tabular}
\tablefoot{The columns show the fractions of contribution from the mass and luminosity of stellar populations that are old ($t>8$ Gyr), of intermediate age ($4<t<8$ Gyr), and young ($t<3$ Gyr). }
\label{tab:mass_light_frac}
\end{table}

Figures \ref{fig:j020536_assembly}, \ref{fig:j205050_assembly}, and \ref{fig:j225546_assembly} illustrate the 1D (right-most panels) and 2D (first and middle panels) SFHs from the mass-weighted and the light-weighted stellar populations in the top and bottom rows, respectively. The 2D SFHs are depicted by the mass or light fractions in the age-[M/H] grid, with the weights coloured according to the colour bar on the right of each plot. In the 1D SFHs, the inner component is shown in red and the outer component in blue. The $x$ axis in both depictions is time, but for the 2D SFH, the time is denoted as the logarithmic stellar age in years, while for the 1D SFH, this is instead denoted by the look-back time in gigayears. For galaxy J225546, there is an additional `component' coloured in grey, which is in fact the neighbouring galaxy described in Sect. \ref{subsec:voronoi}. This is only depicted here to emphasise the reliability of \texttt{BUDDI} in modelling multiple inherent components and external galaxies simultaneously, and it does not play a part in the inferences from the study.

\subsubsection{Mass fractions}
\label{subsubsec: mass_frac}

The weights from \texttt{pPXF} measure the fractions of stellar populations that contribute to the total stellar mass of the galaxy or object of interest. As a preliminary overview, in Table \ref{tab:mass_light_frac}, we compute the mass contributions from stars that are older than 8 Gyr ($t_{old}$), of intermediate ages between 4 and 8 Gyr ($t_{inter})$, and from stars that are younger than 3 Gyr ($t_{young})$. These values are listed for the galaxy overall, and for the inner and outer components. Clearly, the majority of the mass of the ellipticals lies in stars older than 8 Gyr, in each component as well as collectively, making up over 85\% of the mass for all except the outer component of J205050, which only contributes to 64\%. The inner component appears to be entirely composed of such old stars, adding up to 100\% of its mass. The fraction of stars younger than 3 Gyr is null or insignificant, except for the outer component of J225546, where they contribute mildly to 1.5\% of the mass. This small fraction cannot be entirely ruled out as a consequence of the neighbouring object, which might not be completely deblended even when its profile is fitted simultaneously with that of the target galaxy. The stars of intermediate age are prominent only in the outer component, varying from low to substantial in their mass contributions, between $\sim2-35$\%. A more detailed analysis on the mass assembly of the components through cosmic time is described in the following section.

\subsubsection{Star formation histories from mass fractions}

The 2D SFHs built from the mass fractions in general illustrate a picture where the inner component of the elliptical galaxies are extremely old and have higher metallicities. These weights are concentrated in the upper right corner of the grid, with a minute spread in age ($t_{MW}>10$ Gyr), while there is a larger spread in metallicity for galaxies J020536 and J225546, between -0.5 and 0.3 dex. The outer component also shows a dominant old stellar population, with a wider spread in age compared to the inner component, and an even greater spread in metallicity, with stellar populations spanning ages $\sim4-14$ Gyr and the entire metallicity range of the grid. These are indicative of minor episodes of continuous and slower stellar mass assembly that have occurred after the primary peak of star formation, wherein different stellar populations have been accreted.

The 1D mass-weighted SFHs trace the cumulative mass assembly of each component through cosmic time. They complement the results from the 2D SFHs with a clear depiction of the inner component having assembled its stellar mass initially and swiftly within a very short timescale, forming the inner core of the galaxy. The peak mass assembly of the outer component exhibits a delay of several gigayears compared to the inner component, and has assembled its mass at a slower rate, over the course of a few gigayears, most likely growing its mass by continuous accretion during this time until attaining the present-day mass. The SFHs for the inner and outer components exhibit a flat slope from $\sim10$ Gyr and $\sim5$ Gyr ago, respectively, up to $z=0$, marking their quiescence during this period. This clearly points to an inside-out mass growth scenario for local elliptical galaxies. 

Furthermore, the timescales $\tau_{50}$ and $\tau_{90}$ were measured from the cumulative mass fractions (Table \ref{tab:all_stellar_pops}). $\tau_{50}$ measures the look-back time when half of the present-day stellar mass had been assembled, while $\tau_{90}$ measures the same for 90\% of the present-day mass, which is related to the onset of quenching in the galaxy \citep{ferre-mateu2018a, romero-gomez2024}, especially for retired quiescent galaxies. For all three ellipticals, the inner component had already formed half of their mass by $\sim13$ Gyr ago, while the outer component had taken $\sim1-5$ Gyr longer to assemble the same (with timescales of 2.09 Gyr, 4.74 Gyr, and 670 Myr for J020536, J205050, and J225546, respectively). Similarly, the inner component appears to have quenched its star formation activity as early as $\sim12$ Gyr ago. The outer component, on the other hand, stopped forming stars only $\sim6.5-10.5$ Gyr ago, roughly $3-5$ Gyr after half the stellar mass had assembled. The relative contribution of the inner component to the total stellar mass was found to be 50\% for all three galaxies, indicating that half of the stellar mass originates from the outer component. These collective findings demonstrate an early and rapid assembly of the majority of the stellar content within the inner component of the elliptical galaxies, which consequently appears to be a `co-dominant' component in terms of its influence on the present day total stellar mass. The longer assembly timescales attributed to the outer component, along with it hosting half of the galaxy stellar mass, imply a slower addition of stars well after the inner component has quenched its star formation.

\subsubsection{Light fractions}
\label{subsubsec: light_frac}

Similar to the mass fractions described in Sect. \ref{subsubsec: mass_frac}, the light fractions are estimated by \texttt{pPXF}, from which the fraction of luminosity contributions of different stellar populations can be inferred. The luminosity of the galaxies is predominantly composed of an old stellar population ($t_{LW}>8$ Gyr), contributing nearly 100\% of the light on average. This is mirrored in both the inner and outer components as well, with the young stars of $t<3$ Gyr having no contribution to the luminosity. The intermediate-age stars contribute substantially more to the outer components, accounting for $\sim20-37$\%. The exception to this is the outer component of J225546, for which the intermediate stars only accounting for $\lesssim2\%$ of the luminosity of the galaxy. The SFHs of each component are described in depth below, which puts into perspective these contributions to the galactic light across cosmic time. Based on the mass and light fractions of various stellar ages, the evident conclusion is that these ellipticals are indeed substantially evolved galaxies with no recent star formation.

\subsubsection{Star formation histories from light fractions}

The SFHs drawn from the light fractions are shown in the lower panels of Figs. \ref{fig:j020536_assembly}, \ref{fig:j205050_assembly}, and \ref{fig:j225546_assembly}, respectively. The 1D cumulative light-weighted age distributions (right-most panels of the above-mentioned figures) serve as a qualitative proxy for the SFH, reflecting the relative contributions of stars formed at different cosmic epochs, to the current galaxy luminosity. Since the light-weighting is performed with the $V-$band luminosity, the results also trace when star formation episodes had occurred during more recent times. $\tau_{50}$ and $\tau_{90}$ derived from the cumulative light fractions (Table \ref{tab:all_stellar_pops}) indicate the look-back time when the stars contributed to 50\% and 90\%, respectively, of the luminosity of the galaxy or its components. In this case, $\tau_{90}$ does not necessarily highlight the onset of quiescence, but rather highlights the epoch when the stars contributed to 90\% of the present-day luminosity. We find that the light-weighted 2D SFHs follow the trends observed in the mass-weighted SFHs, revealing a similar pattern in the history of star formation. In J205050, a significant portion of the light originates from stellar populations older than 5 Gyr, spanning a broad range of metallicities from super-solar to very sub-solar, reflecting a diverse range of stellar populations, similar to the mass-weighted 2D SFH. In contrast, the outer component of J225546 is dominated primarily by stars older than 10 Gyr, but with a wide spread in metallicities.

In galaxy J020536, the old metal-enhanced stars still dominate the luminosity of the inner component, suggesting that the bulk of the stellar mass assembly corresponds to a single significant star formation event more than 10 Gyr ago, which formed rapidly and very early on in the life of the Universe. This component appears to have formed the majority of its stars that now contribute to  90\% of the luminosity 12 Gyr ago. In the outer component of J020536, there appears a smaller fraction of a younger stellar population aged $\sim4$ Gyr with lower metallicities reaching sub-solar metal abundances. The outer component has taken an additional 5 Gyr to form or assemble the stars dominating 90\% of the light. The 1D SFHs also show a long period of constant cumulative light fraction in the inner component, where it appears to have achieved quiescence and remained undisturbed by new star forming episodes or accretion in the last 4 Gyr.

The light fractions of J205050 similarly show a predominantly old and metal-enhanced light-weighted stellar population in the inner component, which, similar to J020536, has undergone a rapid event of bulk star formation at early epochs. The $\tau_{90}$ look-back time, indicating the major star formation in this component, seems to have occurred and ceased roughly 13.5 Gyr ago, followed by a flat unperturbed quiescent slope leading to the present-day. The outer component, however, shows a delayed contribution to 90\% of the luminosity, with most of these stars having formed around 7 Gyr ago. It has remained quiescent since then, with no significant star formation up to the present day.

Following this trend, the 2D light-weighted SFH of J225546 shows an inner component again dominated by the light of the old metal-rich stars at solar and super-solar metal abundances. The outer component
consists of two distinct light-weighted stellar populations, one with old and metal-enhanced stars, and the other with similarly old but quite metal-poor stars reaching [M/H]$_{LW}<-1$ dex. The cumulative light fractions complement this with a clear indication that the peak of star formation in both the inner and outer components happened at earlier look-back times.  The outer component followed after the principal star formation occurred, but the formation timescale between both components for this galaxy is not as substantial as for the other two galaxies, which is also evident from their $\tau_{90}$, indicating a delay of roughly 3 Gyr.

The stellar population analyses and the reconstructed SFHs support the two-phase scenario introduced in Sect. \ref{sec:intro}. The inner regions of the ellipticals ($R_e<5$ kpc) could, in principle, be associated with the in situ star formation within the galaxy as a result of dissipative collapse or major mergers very early on in their lifetimes. In contrast, the outer component ($6<R_e<20$ kpc in this sample), is likely a combination of star formation formed in situ from gas, with the majority of the stellar envelope having been accreted through dry mergers over longer timescales compared to the inner component. Regarding the former, \citet{choi2024} find that in cosmological simulations, mergers can result in the accretion of gas from external galaxies. They also find signatures of smoothly accreted gas, particularly through the cooling of the galaxy halo, aligning with the findings of \citet{lagos2015} in ETGs. In \citet{choi2024}, another origin is attributed to gas from the primary galaxy halo that has been recycled over generations of stellar evolution. While these processes were examined in the context of gas accretion in the inner regions of galaxies fuelling a supermassive black hole, a smaller fraction of the gas can be distributed over different regions, including the outskirts. On the other hand, the accretion of direct stellar material from dry mergers with external satellites of different metallicities occurs around the old, metal-rich core, which undergoes minimal evolution since its star formation was quenched early on.

\section{Discussion}
\label{sec:discussion}

In this work, we have decomposed three elliptical galaxies of log($M_*/M_\odot) \approx 11$, observed with MUSE, into an inner and outer component using the spectro-photometric code \texttt{BUDDI}. We have analysed the mass-weighted and light-weighted stellar populations of each component, and reconstructed their SFH to study the processes involved in their formation and subsequent quenching. The radial gradients of the stellar population properties were also measured as a complement to these analyses. In the following section, these results will be explored in the context of the formation channels of present-day elliptical galaxies. 

\subsection{Multiple structural components in ellipticals: Implications from surface brightness profiles}

The notion that elliptical galaxies may consist of multiple components has been examined in several studies through 2D photometric decomposition, which are outlined in this section. This work builds upon existing methods by incorporating the imaging and spectroscopic capabilities of IFS.  The spectro-photometric decomposition technique with \texttt{BUDDI} not only models the surface brightness profiles of various components at each wavelength of an IFU datacube, but also enables a more accurate derivation of stellar populations from the resulting spectra compared to photometric colours, which often suffer from the age-metallicity degeneracy \citep{worthey1999}.

\citet{huang2013b} investigated the possibility of multi-component ellipticals in a sample of 94 local galaxies in the CGS survey. The high spatial resolution of deep optical images from this survey allowed for a reliable 2D decomposition of the ellipticals into three to four sub-components using \texttt{GALFIT} \citep{peng2002galfit, peng2011galfit}, so that they could study their structural properties in the $V$ band. They found that for the majority of their sample the surface brightness profiles were best modelled by three Sérsic components, representing a compact inner core, a middle component, and an outer extended envelope. The 2D residuals consistently improved with increasing complexity in the models. Interestingly, they reported only a small fraction of their sample to be modelled with two components. Furthermore, they also suggested that their high-luminosity sample of ellipticals showed photometric signatures of being `core' ellipticals with a shallow slope within the innermost component. \citet{lacerna2016} report similar findings, noting a good fit with three Sérsic models for a sample of 89 isolated local elliptical galaxies from SDSS images. \cite{huang2013b} and \cite{huang2013a} linked the inner and intermediate components to the phase of elliptical galaxy formation involved in situ dissipative processes, while associating the outer component with a later stage of ex situ star accretion through dry mergers. These findings and interpretations closely align with the ones presented in this work, despite differences in the number of components modelled.

Along these lines, \citet{spavone2017} performed a detailed analysis of the surface brightness profiles of six massive ETGs in A VST Early-type GAlaxy Survey (VEGAS) survey \citep{capaccioli2015}. They adopted various profile combinations to identify the most accurate model. Four galaxies were effectively modelled using a Sérsic + exponential profile, whereas the remaining two were better represented by a double Sérsic model. Additionally, they found that a model consisting of two Sérsic profiles for the inner regions and an exponential profile for the outer region generally provides a good fit for all galaxies; this model is the primary one they used for the analysis. Following theoretical predictions, they conclude that the inner Sérsic profile represents a component formed by in situ processes. The middle Sérsic profile represents the dominant component composed of a phase-mixed accreted component where the material has dynamically relaxed. The outer exponential profile is then associated with the unrelaxed accreted component, which often shows low-surface-brightness signatures of interactions, like shells or streams. From the 1D surface brightness profiles, they located inflections that are indicative of physical transitions between the different components. A brief comparison with our 1D surface brightness profiles reveals that the inflection occurs at a galactocentric distance of $\sim3.5-4$ kpc for the three ellipticals. The sizes of the inner in situ component from our models lie within this apparent transition radius with $R_e \sim 1.5-3$ kpc. However, given the different model combinations used in both studies and the fact that our 1D surface brightness profiles are constructed solely from the $r-$band image slice of the datacube, we cannot conclusively determine if the inner component corresponds to the transition radius observed in the surface brightness profiles, as this radius can shift depending on the wavelength employed. Despite the distinct methodologies, both the photometric and spectro-photometric models support the idea that the inner component is primarily shaped by in situ star formation, while the outer component is dominated by ex situ accretion, which is consistent with theoretical predictions and simulations.

\subsection{Stellar population properties: Inside-out formation and the two-phase scenario}
\label{subsec:discussion_SP}

Several studies in the literature have investigated the mode of mass assembly in elliptical galaxies using IFS. However, most of these studies address this question by measuring the radial gradients of stellar age and metallicity. Our methodology instead measures the mean stellar population properties in each of the two `clean' components extracted with the spectro-photometric code \texttt{BUDDI}, where the mixing of light is mitigated between the components. Therefore, the comparisons that we make to literature in this section are less specific to the measurements and techniques, and rather focus on the general results and implications to elliptical galaxy formation.

With the diverse range of methods and observations used to tackle this, there exists a broad variation in the stellar population results reported by different IFS-based studies. For instance, \citet{gonzalez-delgado2015} report elliptical galaxies in the Calar Alto Legacy Integral Field Area Survey (CALIFA) survey showing slight negative gradients in light-weighted age, with $\sim10$ Gyr in the centre and dropping to $\sim5$ Gyr at 1$R_e$. A notable outcome by \citet{zibetti2020} in their light-weighted stellar age analysis for 48 elliptical galaxies in CALIFA was the U-shaped profile for the ages, with the transition occurring at $\sim0.4 R_e$. Otherwise, in contrast to the other studies, they find a slightly positive gradient up to $1.5 R_e$, beyond which the slope flattened out. The authors attribute this inversion to the interplay between competing quenching mechanisms: the outside-in quenching due to feedback-driven winds is favoured at large radial distances, while inside-out quenching triggered by AGN-driven feedback occurs starting from the galactic centre moving outwards. Moreover, they suggest that these results lend support to the two-phase formation scenario, wherein the quenching mechanisms likely compete during the initial phase. The flattening of the slope at the outermost radial regions could then be attributed to the accretion of ex situ stars from satellite galaxies, where their effect is the most prominent. With the Mapping Nearby Galaxies at APO (MaNGA) survey, \citet{lacerna2020} studied the stellar population properties of classical ellipticals, recently quenched ellipticals, and blue star-forming ellipticals independently. Our sample of ellipticals fall under the classical category, and we limit our comparisons to this sub-sample. They found moderately negative gradients in the light-weighted ages, and flat gradients in the mass-weighted ages. Similar flat age gradients were reported in \citet{parikh2019}, while weak negative age gradients were found in  \citet{dominguez-sanchez2019} for elliptical galaxies in MaNGA, and similarly in \citet{riffel2023} for ETGs with MaNGA. A different study with MaNGA by \citet{goddard2017} reported contrasting slightly positive gradients for ETGs, which include both ellipticals and S0s.
    
In principle, from these studies, a positive gradient in stellar age implies the galaxy has assembled its mass from the outside in. A negative gradient, on the other hand, suggests an inside-out mass growth, starting from the innermost regions of the present-day observed elliptical galaxy. Along those lines, the flat gradients observed in some works would indicate a more uniform mass growth throughout the galaxy. However, the stellar populations of galaxies, like most of their properties, are driven by a multitude of processes that are more likely to be interwoven rather than relying on a single mechanism. Our results from the stellar population analysis fall within the wide diversity of results that have been observed in the studies mentioned above. Both the mean mass-weighted and light-weighted stellar age differences between the inner and outer components were low to moderate ($\sim1-5$ Gyr). For the three elliptical galaxies in this sample, this observation insinuates a mild inside-out formation scenario for J020536 and J225546, and a more pronounced inside-out formation for J205050. This is qualitatively and globally similar to the moderately negative or flat gradients reported in a few of the aforementioned IFS-based studies. In the context of a two-component galaxy, the inner component would have formed stars earlier, while the star formation continued for a slightly longer period, building up the stellar mass on the outskirts. Such old stellar populations are typical for galaxies that have an early intense star formation in the inner component occurring in an extremely short timescale, which quench soon after a few gigayears. The younger ages in the outer component could then be associated with accreted stars from satellite galaxies that are themselves quenched and are nearly as old as the main elliptical galaxy. At this point, we note again the high levels of degeneracy associated with stellar populations older than 8 Gyr, and it is important to be cautious when interpreting such old ages.

While the mean stellar ages offer some perspective into the growth mode of these galaxies, they are solely not enough to confirm this theory. The list of literature results described earlier in this section additionally report results on the stellar metallicity analyses for their samples of ellipticals and ETGs. Most of the work done along these lines in the past with IFS have consistently reported negative metallicity gradients, with only the slope signifying their steepness varying between them. The exception is \citet{gonzalez-delgado2015}, where they find nearly flat gradients in mass-weighted and light-weighted stellar metallicities for an elliptical galaxy sample in CALIFA. The reason for this discrepancy is not fully understood, and could be linked to the different methodologies employed in different studies. For an extensively shared sample, \citet{zibetti2020}, on the other hand, notice steeper gradients up to $1.5 R_e$, after which the profiles transition to a flatter slope. Consistent with this, for their sample of classical ellipticals, \citet{lacerna2020} find mildly steep gradients for the mass-weighted and light-weighted metallicities up to $1 R_e$, beyond which the profiles flatten. Negative metallicity gradients are in accordance with the results presented \citet{parikh2019}, \citet{dominguez-sanchez2019}, and \citet{goddard2017}. 

Based on these studies on stellar metallicity gradients, a negative slope points to one possible scenario of early dissipative collapse from a gas-rich cloud already enhanced with metals. The formation and metallicity profiles of ellipticals were investigated via simulations in \citet{kobayashi2004}, where the relevant physical processes including radiative cooling, stellar feedback through SN and winds, and chemical enrichment models were incorporated. However, the AGN feedback often observed in massive galaxies was not considered. Under this simulated regime, they find that the radial metallicity gradients trace SFHs, and require a combination of both rapid dissipative collapse and major mergers to adequately explain their variation. Moreover, they expect that subsequent major mergers tend to taper out the radial metallicity profiles, diminishing signatures of physical processes prior to the merger. If the progenitor mass ratios are high enough, the metallicity gradients can be completely destroyed and would appear flat in present-day observations. Additionally, the variation in metallicity gradients can be triggered by late gas accretion in the outer regions, where a slower phase of star formation is then induced. With the inclusion of AGN feedback in \citet{taylor2017}, they reinforce the flattening of metallicity gradients by major mergers in massive galaxies, but the quenching induced by AGN makes it challenging to regenerate the gradients through later in situ central star formation.

Our results show a substantial decrease in the metal enhancement for the outer component, compared to the metal-rich inner component, observed in both the mass-weighted and light-weighted properties. A clear negative radial gradient of metallicity is seen qualitatively in the 2D maps as well, with a decline from the central region of the galaxies. This points to an intense in situ star formation phase in the inner component, which might have been influenced by major mergers. This is difficult to discern from our analysis since our estimates depict the mean stellar metallicity for each component, and therefore any information on gradients that might or might not exist within the inner component is lost. The decline in metallicity for the outer component, in tandem with the old stellar ages presented by these stellar populations, can be explained on the basis of accretion of ex situ stars from old satellite galaxies that have formed their stars from metal-poor gas. In a dry merger, the ex situ stars are deposited on the outskirts of the galaxy due to its lower binding energy compared to the in situ formed stars \citep{spavone2017}, causing accreted stars to dominate in the outer component. The combined stellar population properties presented in this work are therefore compatible with the two-phase scenario described in Sect. \ref{sec:intro}, with a preferred inside-out mass growth for the intermediate-mass ellipticals comprising our sample.

\subsection{Star formation histories: Stellar assembly across time }

The star formation of ETGs has been extensively investigated in several works with IFS surveys, using a variety of techniques, from which some of the most relevant ones will be described in this section. Similar to the stellar population gradients, our approach allows us to analyse the SFH of the inner and outer components of each elliptical galaxy, identifying the epochs of extensive star formation and quenching. These results from the literature, including ours from this study, align well with the two-phase scenario predicted in simulations of spheroid-dominated galaxies \citep{naab2009, oser2010, johansson2012}.

Even from a cursory look, it would appear that the ellipticals in our sample fit the description of red and dead retired galaxies commonly referenced in the literature. This impression is supported by the cumulative SFHs (Fig. \ref{fig:j020536_assembly}, \ref{fig:j205050_assembly}, and \ref{fig:j225546_assembly}), which reveal a significant early and rapid assembly of stellar mass for the inner component, heavily influencing its mass and luminosity. Similarly, the outer component exhibits a concentrated period of star formation in its early life, resulting in the majority of its stellar mass and luminosity. We surmise that the inner and outer components represent the in situ and ex situ components, respectively, motivated by similar works described in \citet{huang2013a, huang2013b, spavone2017, spavone2021}. Following the initial burst of star formation, the mass assembly histories of both components allude to rather quiescent ellipticals where there has been no significant build-up of stellar mass over roughly the last 8 Gyr. The SFHs that account for the luminosity contributions by the stellar populations in each component depict a slightly nuanced picture. The majority of the light does indeed originate from the initial rapid star formation episode shortly after the Big Bang, and we find no indications of younger stellar populations contributing to the stellar mass or luminosity of any of the galaxies.

The SFHs of the outer components are more extended than those of the inner components, with the peak of star formation and mass assembly shifted towards relatively recent epochs. However, from Table \ref{tab:mass_light_frac}, for galaxies J020536 and J205050, the contribution of stars younger than 3 Gyr to the stellar mass or luminosity is still entirely negligible. The fraction of intermediate-age stars is however quite substantial next to the oldest stars, suggesting this population either formed in situ later in time or was accreted as an ex situ population through dry mergers with younger satellite galaxies aged between 4 and 8 Gyr old. Such a blend of old and intermediate-age stars were measured for dwarf elliptical galaxies in the Virgo and Fornax galaxy clusters \citep{geha2003, michielsen2008}, and in the Local Group \citep{geha2015}. The unperturbed nature of the quiescence of these elliptical galaxies provide additional evidence that they are classical red and dead objects in the local Universe, where both in situ and ex situ star formation have ceased several gigayears ago. In that vein, the following segments of the discussion outline other studies in the literature that have investigated the assembly histories of elliptical galaxies through various observations and techniques.

\citet{lacerna2020} analysed the integrated SFHs of 251 classical elliptical galaxies in MaNGA. They find that these galaxies had built up 70\% of their stellar masses between 8 and 11 Gyr ago, and 90\% of their masses by 5 Gyr ago at the latest. In conjunction with their results on the stellar population gradients (we refer the reader to Sect. ~\ref{subsec:discussion_SP}), they predict that in the earliest stages of elliptical galaxy formation, star formation and the resulting stellar populations were relatively uniform across both the inner and outer regions. It was only in subsequent epochs that the outer regions started forming their stars after the inner regions did. Moreover, by analysing the specific star formation rates of the galaxies, they conclude that, following an initial phase of substantial homogeneous star formation, classical ellipticals continue to grow from the inside out and are subsequently quenched in the same manner. This was naturally explained on the basis of an intense burst of significant star formation, followed by rapid gas consumption, leading to a swift quenching of star formation. Nevertheless, they recognise the role of dry mergers where ex situ stars can be accreted on to the main elliptical, and continue its stellar mass assembly. 
Despite the methodology and sample used being entirely different to theirs, the inferences we present are remarkably similar in the context of the two-phase formation scenario for ETGs. However, it is important to note that the fossil-record method does not directly characterise a given stellar population as having formed via in situ or ex situ processes.

A study by \citet{gonzalez-delgado2017} employed the fossil-record method to reconstruct the spatially resolved 2D SFHs of galaxies spanning a wide variety of morphologies in the CALIFA survey. Similar to Table \ref{tab:mass_light_frac} in this work, they report the fractions of the stars in different stellar age bins that contribute to the luminosity of the galaxy. They find that $\sim3$\% of stars younger than 1 Gyr contribute to the light from the outer regions of ellipticals, while the older stars with ages over 4 Gyr dominate with $\sim86$\% of the populations. In the inner regions, that fraction increases to 96\% for the older stars, and 1\% for the younger population. We note that this is slightly in contrast to our sample, which reveals no traces of young stellar populations in the inner components. The contributions to the stellar mass follow the same pattern in the inner and outer regions. Furthermore, they find evidence of an extended phase of star formation since 4 Gyr ago, and signs of the outer regions of ellipticals that are actively forming stars after the quiescent phase.  

Another sample of nearby ETGs from the $\mathrm{ATLAS^{3D}}$ survey \citep{cappellari2011atlas} were investigated in \citet{mcdermid2015atlas}, where they measure the spatially resolved stellar populations and integrated SFHs within $1 R_e$ of the galaxies. Assessing this region within $1 R_e$ across the galaxies in our sample (see Table \ref{tab:params}), we note that this would result in a loss of information about the outskirts or the outer component, which extend to at least twice the $R_e$ from the single Sérsic model. They find that roughly half the present-day stellar mass was already assembled as early as 2 Gyr after the Big Bang. The most massive present-day ETGs with $M_*>10^{10.5}M_\odot$ host these old stars, and had built up 90\% of the stellar mass by $z\sim2$, which corresponds to a look-back time of approximately 10.5 Gyr ago under the cosmology defined in Sect. \ref{sec:intro}. They attribute their results to the picture where the old and metal-rich ETGs were formed from an early dissipative process at high redshifts, which is followed by a phase of quenching. Subsequently, stellar accretion takes over as the dominant mechanism of stellar mass growth, aligning with the two-phase scenario of formation.

The importance of ex situ stars that have been accreted by ETGs in their later stages has been demonstrated by observations and simulations, particularly with the IFS surveys. The metallicity and age distributions and profiles are often exploited to estimate the fraction of ex situ accreted stars in an ETG. For instance, \citet{oyarzun2019} measure the ex situ stellar mass fractions of ETGs in the MaNGA survey. They find that for ETGs of $M_*\gtrapprox10^{11.5}M_\odot$, ex situ stars make up to $80\%$ of their stellar mass, at a galactocentric distance of $\sim2 R_e$. Similarly, using a combination of full-spectral fitting and galaxy evolution models, \citet{davison2021} identify ex situ stars in ETGs observed with MUSE, and report that the fraction of accreted stars increases with galactocentric radius, with values ranging between $\sim30$ and 100\% at $\sim2 R_e$. 
\citet{spavone2021} present deep imaging of three massive galaxies from the VEGAS survey, combined with MUSE observations. This study correlates the structural components from imaging with the kinematics and stellar populations from IFS analysis, to explore the mass assembly histories of these galaxies. They find ex situ mass fractions of 77\%, 86\%, and 89\%, indicating that accretion is the primary contributor to the stellar mass. Our results align with the ranges reported by some of these studies, indicating that 50\% of the stellar mass was accreted from external galaxies, forming a second co-dominant component in the context of a dual-component picture of elliptical galaxies.

\section{Summary and conclusions}
\label{sec:conclusions}

In summary, this study was primarily focused on the identification of the different components present in elliptical galaxies of intermediate stellar mass in the local Universe. For a sample of three galaxies observed with MUSE, we successfully modelled the surface brightness profiles adopting two Sérsic components, representing an inner and an outer component, using the spectro-photometric decomposition code \texttt{BUDDI}. After this, the full-spectral fitting software \texttt{pPXF} was used to fit the cleanly extracted spectra and estimate the mass-weighted and light-weighted stellar populations of each component, and subsequently their SFHs. Spatially resolved 2D stellar population maps were constructed after binning the datacubes to achieve a sufficient minimum S/N threshold using the Voronoi tessellation method. We investigated the physical properties of each component to constrain the formation scenarios and the mode of stellar mass growth in local elliptical galaxies. While similar studies with photometric decomposition on data from imaging surveys have been conducted, this approach of using IFS to model multiple components in datacubes is relatively new and offers a different perspective of exploring elliptical galaxy formation and evolution. The main conclusions of our study are listed below:

\begin{itemize}

    \item The best fit for the surface brightness profiles of the elliptical galaxies in this sample was achieved using two Sérsic components with distinct properties, referred to as the inner and outer components.

    \item The stellar population analysis of the inner and outer components and the spatial maps reveal an inside-out formation and growth mode in ellipticals. The majority of the stellar populations are older than 8 Gyr for both components (100\% for the inner component, and $\gtrsim65\%$ for the outer component), from both the mass and light-weighted ages. The inner components are generally more metal-enhanced than the outer ones. 

    \item The reconstructed SFHs highlight an inner component that has formed and assembled the bulk of its stars in the earliest epochs of the Universe, before $z\sim2$. This component is expected to have formed from interlinked physical processes: rapid dissipative collapse and major mergers, forming a central core hosting the oldest stellar populations. The outer component, however, aligns more with accretion events from past dry mergers, in which old metal-poor stars have been deposited on the outskirts. This process continues to build the stellar mass of the ellipticals without having to undergo a new phase of star formation. The fraction of mass and luminosity contributed by the different stellar populations in the two components further confirms the two-phase assembly mode.

    \item The mass weights from \texttt{pPXF} additionally provide information about the formation ($\tau_{50})$ and quenching ($\tau_{90})$ look-back times, which show that half of the present-day stellar mass in our sample had already assembled $\sim12$ Gyr ago, and had undergone quenching around $\sim5$ Gyr ago at the latest. The light-weighted properties show similar trends, except for $\tau_{90}$, which indicates a more extended SFH in the outer component, with 90\% of the light contribution coming from look-back times as recent as $\sim5$ Gyr ago.

\end{itemize}

In conclusion, our method of decomposing the multiple components of elliptical galaxies has helped trace back the signatures of the physical processes involved in their formation and evolution across cosmic time. This serves as an initial pilot study, which can be extended to a more statistical sample such as the ellipticals in BUDDI-MaNGA \citep[e.g.][]{jegatheesan2024} to better constrain the formation pathways. Possible future follow-up work involves capturing the extremely low-surface-brightness stellar halos of elliptical galaxies and modelling their components \citep{johnston2018}. Additionally, studying a more representative sample of local ellipticals, encompassing a range of masses from dwarf to giant ellipticals, variations in star formation activity, and different environments, would provide a comprehensive understanding of the formation and evolution of these galaxies.

\begin{acknowledgements}

KJ acknowledges financial support from ANID Doctorado Nacional
2021 project number 21211770. KJ thanks the LSST-DA Data Science Fellowship Program, which is funded by LSST-DA, the Brinson Foundation, and the Moore Foundation; her participation in the program has benefited this work. EJJ acknowledges support from FONDECYT Iniciaci\'on en investigaci\'on 2020 Project 11200263 and the ANID BASAL project FB210003. AEL acknowledges the support from Coordena\c{c}\~ao de Aperfei\c{c}oamento de Pessoal de N\'{i}vel Superior (CAPES) in the scope of CAPES-PROEX fellowship, process number 88887.513351/2020-00, as well as Conselho Nacional de Desenvolvimento Científico e Tecnológico (CNPq). RR acknowledges support from  Conselho Nacional de Desenvolvimento Cient\'{i}fico e Tecnol\'ogico  (CNPq, Proj. 311223/2020-6,  304927/2017-1, 400352/2016-8, and  404238/2021-1), Funda\c{c}\~ao de amparo \`{a} pesquisa do Rio Grande do Sul (FAPERGS, Proj. 19/1750-2 and 24/2551-0001282-6) and Coordena\c{c}\~ao de Aperfei\c{c}oamento de Pessoal de N\'{i}vel Superior (CAPES, Proj. 0001).
ACS acknowledges support from CNPq, Proj 314301/2021-6 and CAPES, Proj. 0001.   
This work was based on observations collected at the European Organisation for Astronomical Research in the Southern Hemisphere under the ESO programme 099.B-0411 (PI. Johnston). 

This work made use of \texttt{Astropy} (\url{http://www.astropy.org}), a community-developed core Python package and an ecosystem of tools and resources for astronomy \citep{astropy:2013, astropy:2018, astropy:2022}, \texttt{MPDAF }\citep[\url{https://mpdaf.readthedocs.io/}; ][]{mpdaf}, and the \texttt{ESO SkyCalc} Web Application. The authors thank Amy Jones for useful discussions related to \texttt{SkyCalc} usage.

\end{acknowledgements}

\bibliographystyle{aa} 
\bibliography{papers} 

\begin{appendix}

\onecolumn
\section{Decomposed spectra from \texttt{BUDDI}}

This appendix shows the observed spectra and the decomposed spectra from \texttt{BUDDI} for galaxies J205050 and J225546. This is meant to demonstrate the reliability of the decomposition technique and that the resulting spectra are visually robust for all three galaxies.  

\begin{figure*}[h!]
    \centering {\includegraphics[trim=2.8cm 0.4cm 1.8cm 1.2cm, clip, width=0.92\textwidth] {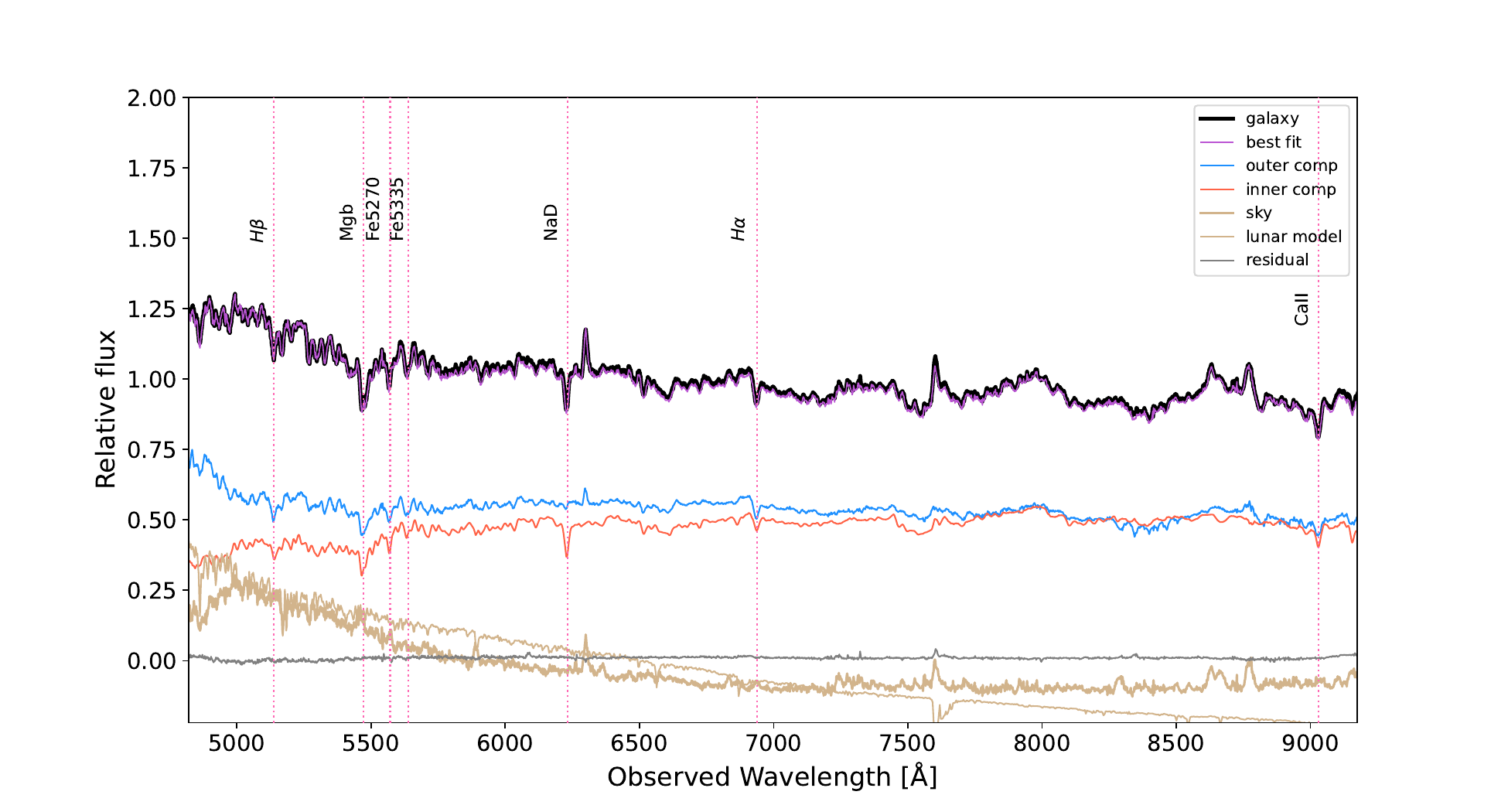}}
    \caption{The decomposed spectra of the different components in J205050 extracted by \texttt{BUDDI}. The colour scheme is the same as Fig. \ref{fig:decomposed_J020536}. This spectrum has a strong contribution from the scattered moonlight, and reasonably matches the shape of the sky spectrum obtained with \texttt{BUDDI}.}
    \label{fig:decomposed_J205050}
\end{figure*}

\begin{figure*}[h!]
    \centering
    \includegraphics[trim=2.8cm 0.4cm 1.8cm 1.2cm, clip, width=0.92\textwidth]{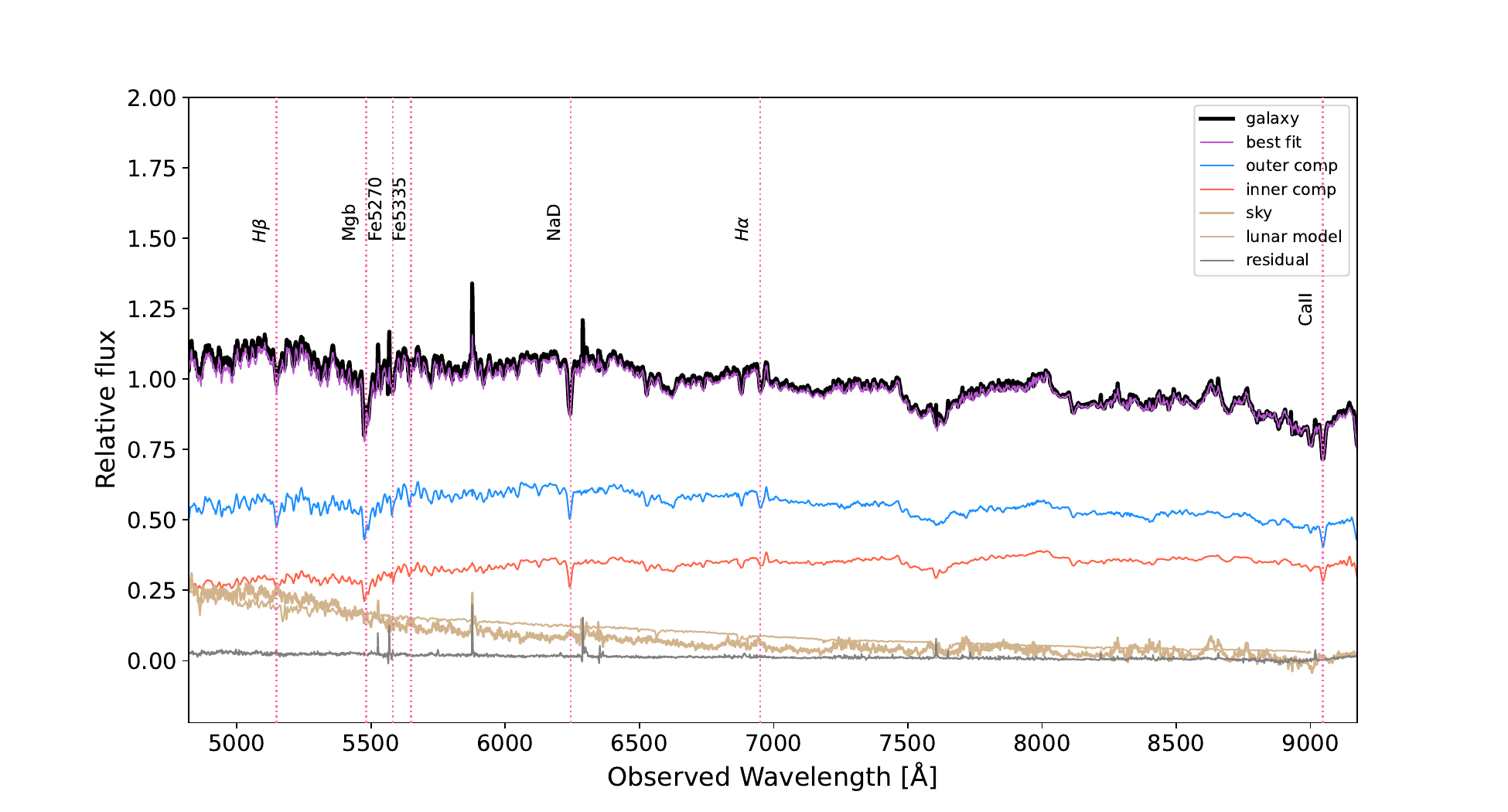}
    \caption{The decomposed spectra of the different components in J225546 extracted by \texttt{BUDDI}. The colour scheme is the same as Fig. \ref{fig:decomposed_J020536}. This galaxy spectrum also has a strong contribution from the scattered moonlight, similar to Fig. \ref{fig:decomposed_J205050}, and reasonably matches the shape of the sky spectrum obtained with \texttt{BUDDI}.}
    \label{fig:decomposed_J225546}
        
\end{figure*}

\end{appendix}

\end{document}